\documentclass[aps,reprint,rmp,letterpaper,unsortedaddress,longbibliography]{revtex4-1}
\usepackage{graphicx}
\begin{document}

\title{Stimulating Uncertainty: Amplifying the Quantum Vacuum with Superconducting Circuits} 

\author{P. D. Nation}
\affiliation{Advanced Science Institute, RIKEN, Wako-shi, Saitama, 351-0198 Japan}
\affiliation{Department of Physics, University of Michigan, Ann Arbor, Michigan 48109-1040 USA}
\author{J. R. Johansson}
\affiliation{Advanced Science Institute, RIKEN, Wako-shi, Saitama, 351-0198 Japan}
\author{M. P. Blencowe}
\affiliation{Department of Physics and Astronomy, Dartmouth College, Hanover, New Hampshire 03755-3528 USA}
\author{Franco Nori}
\affiliation{Advanced Science Institute, RIKEN, Wako-shi, Saitama, 351-0198 Japan}
\affiliation{Department of Physics, University of Michigan, Ann Arbor, Michigan 48109-1040 USA}

\begin{abstract}
The ability to generate particles from the quantum vacuum is one of the most profound consequences of Heisenberg's uncertainty principle.  Although the significance of vacuum fluctuations can be seen throughout physics, the experimental realization of vacuum amplification effects has until now been limited to a few cases.  Superconducting circuit devices, driven by the goal to achieve a viable quantum computer, have been used in the experimental demonstration of the dynamical Casimir effect, and may soon be able to realize the elusive verification of analogue Hawking radiation.  This article describes several mechanisms for generating photons from the quantum vacuum and emphasizes their connection to the well-known parametric amplifier from quantum optics.  Discussed in detail is the possible realization of each mechanism, or its analogue, in superconducting circuit systems.  The ability to selectively engineer these circuit devices highlights the relationship between the various amplification mechanisms.
\end{abstract}

\date{\today}
\maketitle
\tableofcontents

\section{Introduction}\label{sec:intro}
One of the profound consequences of quantum mechanics is that something \textit{can} come from nothing.  Enforced by the uncertainty principle, the vacuum state of quantum mechanics is teeming with activity. Quantum fluctuations inherent in the vacuum give rise to a host of particles that seemingly move in and out of existence in the blink of an eye.  These fluctuations, however fleeting, are the origin of some of the most important physical processes in the universe.  From the Lamb shift \cite{lamb:1947} and Casimir force \cite{casimir:1948,lamoreaux:2007}, all the way up to the origin of the large scale structure \cite{springel:2006} and the cosmological constant \cite{weinberg:1989} of our universe, the effects of the quantum vacuum permeate all of physics.
  
Although the significance of vacuum fluctuations has been appreciated since the early days of quantum mechanics [see, e.g., \cite{milonni:1993}], the quantum properties of the vacuum state constitute an area of quantum field theory that remains relatively unexplored experimentally.  So far, static quantum vacuum effects such as the Casimir force \cite{lamoreaux:1997} and Lamb shift \cite{lamb:1947} have been verified experimentally, along with the recent demonstration of the dynamical Casimir effect \cite{moore:1970,lahteenmaki:2011,wilson:2011}.  In contrast, other dynamical amplification mechanisms such as the Schwinger process \cite{schwinger:1951}, Unruh effect \cite{unruh:1976}, and Hawking radiation \cite{hawking:1974,hawking:1975}, have yet to been observed\footnote{As discussed in Sec.~\ref{sec:analogue-hawking}, recent experimental evidence for an analogue of Hawking radiation \cite{belgiorno:2010} does not go far enough to definitively confirm the existence of this effect.}. The difficulties in observation can be traced to the extreme conditions under which these dynamical phenomena become appreciable.  For example, the dynamical Casimir effect requires rapidly modulating the boundary conditions of the electromagnetic field, with peak velocities close to the speed of light.  Likewise, Hawking radiation not only requires a black hole, but also demands one with a sufficiently small mass so as to make the emitted radiation observable above the ambient cosmic microwave background.  With difficulties such as these in mind, researchers have looked to analogue systems that are able to generate the desired amplification effects, and at the same time surmount the difficulties inherent in observations of the actual processes.  

One such class of available systems are superconducting circuit devices. The quantum mechanics of superconducting circuits has received considerable attention during recent years.  This interest has largely been due to research on quantum computation and information processing \cite{nielson:2000}, for which superconducting circuits \cite{makhlin:2001,you:2005,wendin:2006,clarke:2008,schoelkopf:2008,you:2011} are considered promising fundamental building blocks.  Experimental progress on superconducting resonator-qubit systems \cite{dicarlo:2010} have also inspired theoretical and experimental investigations of quantum optics in the microwave regime \cite{chiorescu:2004,wallraff:2004,schuster:2007,houck:2007,hofheinz:2009}.  These recent advances in the engineering and control of quantum fields in superconducting circuits have also opened up the possibility to explore quantum vacuum effects with these devices.  Indeed, the demonstration of both the Lamb shift in a superconducting artificial atom \cite{fragner:2008}, and the dynamical Casimir effect in a superconducting waveguide \cite{lahteenmaki:2011,wilson:2011}, have already been achieved.

We have two goals in mind for this Colloquium: the first is to introduce to condensed-matter physicists the following quantum vacuum amplification mechanisms: the Unruh effect \cite{unruh:1981}, Hawking radiation \cite{hawking:1974}, and the dynamical Casimir effect \cite{moore:1970,fulling:1976}.  We shall in particular highlight their relationship to the well-known parametric amplifier from quantum optics.  Parametric amplification has been applied extensively in quantum optics to, for example, the generation of nonclassical states \cite{slusher:1985,breitenbach:1997}, tests of wave-particle duality \cite{hong:1987}, quantum-erasers \cite{zou:1991}, and quantum teleportation \cite{bouwmeester:1997,furusawa:1998,kim:2001}. Here we will focus on the physical rather than mathematical aspects of these amplification mechanisms, as others have covered the latter in great detail \cite{birrell:1982,crispino:2008,fabbri:2005,dodonov:2002}.  Our second goal is to introduce to researchers in the high-energy and general relativity communities, possible analogue experimental realizations of these effects in microwave superconducting circuit devices, where the similarities and differences in the various amplification effects manifest themselves in the design of their circuit counterparts.  We emphasize, in particular, the potential advantages arising from their inherently low-noise quantum coherent nature.  

The outline of this Colloquium is as follows:  In Sec.~\ref{sec:amp-intro} we give a brief overview of quantum amplification basics, introducing the formalism to be used in later sections.  Sec.~\ref{sec:vacuum} describes the methods by which photons may be generated from amplified vacuum fluctuations, and highlights the connections between the various effects.  Sec.~\ref{sec:sc-circuits} details the superconducting circuit implementations, as well as reviews progress towards the detection of single-microwave photons, necessary to verify photon production from the vacuum.  Finally, in Sec.~\ref{sec:future} we summarize and briefly discuss possible future applications of superconducting circuit models for engineering quantum ground states and realizing quantum gravity inspired analogues.

\section{Prelude to quantum amplification}\label{sec:amp-intro}
\begin{figure}[t]\begin{center}
\includegraphics[width=8.0cm]{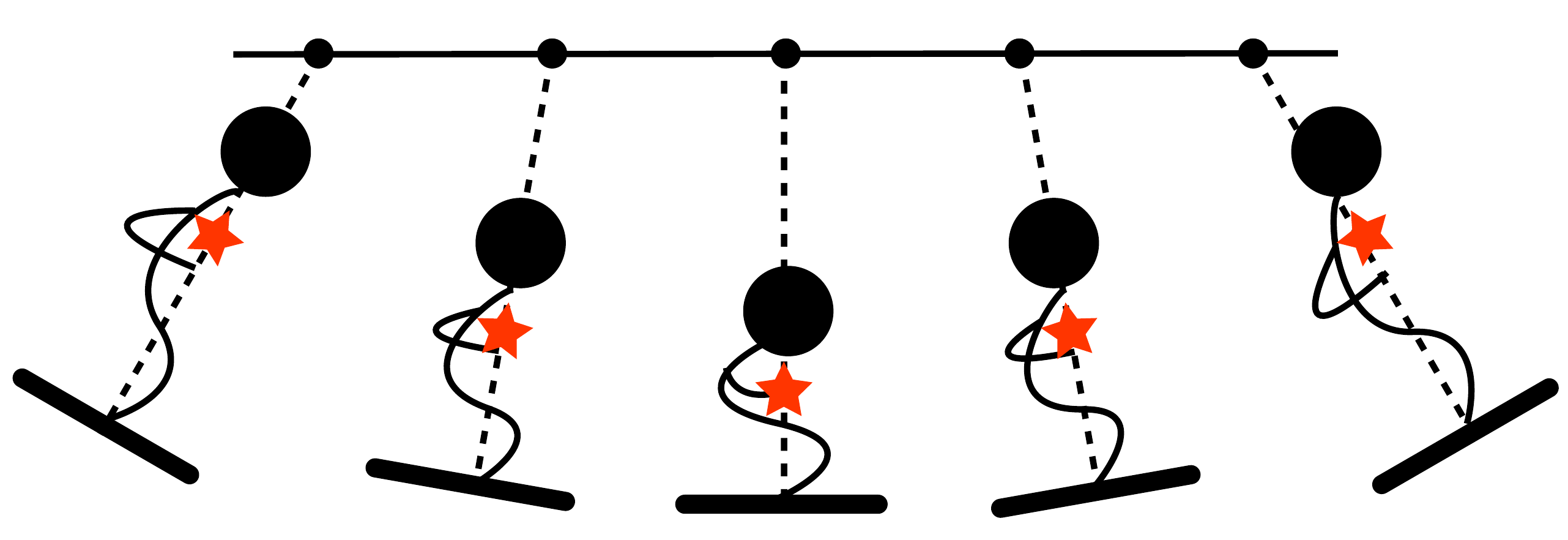}
\caption{(Color online) Parametric amplification of pendulum motion by a child standing on a swing.  The amplification is driven by changing the center of mass (star), and thus effective length, of the pendulum at twice the frequency of the unperturbed swing.}
\label{fig:swing}
\end{center}
\end{figure}

A physical system with time-dependent parameters often has resonant responses at certain modulation frequencies.  This parametric resonance is very general, occurring in a wide variety of both classical and quantum mechanical systems.  The representative example of classical parametric resonance is a child standing on a swing, who periodically modulates her center of mass (CM) by bending at the knees\footnote{Another commonly used example is that of a child  swinging their legs while sitting on a swing.  Careful inspection of the motion however reveals that the child drives the swing at the same frequency as the swing itself.  This situation is therefore better characterized as a driven oscillator rather than a parametric process \cite{case:1990}.}.  For a fixed CM, the equation of motion (for small amplitudes) is that of a simple pendulum with the solution
\begin{equation}
\theta(t)=\theta(0)\cos(\omega_{s}t)+\frac{L(0)}{m\omega_{s}l}\sin(\omega_{s}t),
\end{equation}  
where $L(0)$ is the initial angular momentum and $\theta(t)$ the angular
displacement, while $m$ and $l$ are the pendulum mass and length, respectively. 
With the CM governing the effective length of the swing, this motion modulates
the swing frequency $\omega_{s}=\sqrt{g/l}$ as
$\omega_{s}(t)=\omega_{s}(0)+\epsilon\sin\left(\omega_{\mathrm{cm}}t\right)$,
where $\omega_{s}(0)$ is the unperturbed swing frequency, $\omega_{\mathrm{cm}}$
is the CM modulation frequency, and $\epsilon$ is the resulting small frequency
change in the pendulum motion.  If the child modulates the CM at twice the
oscillation frequency, $\omega_{\mathrm{cm}}=2\omega_{s}$, as shown in
Fig.~\ref{fig:swing},
then the solution to the equation of motion is
\begin{equation}\label{eq:para-swing}
\theta(t)=\theta(0)e^{\epsilon t/2}\cos(\omega_{s}t)+\frac{L(0)}{m\omega_{s}l}e^{-\epsilon t/2}\sin(\omega_{s}t).
\end{equation}
The initial amplitude is therefore exponentially amplified while the out-of-phase component of motion is exponentially suppressed.

For parametric amplification to occur in a classical system it must initially be displaced from the equilibrium state.  This is easily seen by setting $\theta(0)=L(0)=0$ in Eq.~(\ref{eq:para-swing}).  Although many sources of fluctuations can exist, in principle nothing in classical mechanics prevents simultaneously setting the position and momentum of the oscillator to zero.  This is in sharp contrast to the quantum mechanical description of an oscillator where the non-vanishing canonical commutation relation $\left[x,p\right]=i\hbar$ prevents the absence of motion.  This implies that even the ground state of the quantized oscillator contains quantum fluctuations and thus may be parametrically amplified.  The amplification of quantum fluctuations by parametrically modulating the frequency of an harmonic oscillator is closely related to the process of particle production in quantum fields and therefore serves as an instructive example.  We will therefore begin with a short review introducing the basic mathematics and terminology used in later sections by considering the amplification of a quantized oscillator through a time-varying frequency.

We follow the analysis in \cite{jacobson:2004} and begin with the harmonic oscillator described by the Hamiltonian $H=p^{2}/(2m)+m\omega^{2}x^{2}/2$.  With the position and momentum operators obeying the canonical commutation relation $\left[x,p\right]=m\left[x,\dot{x}\right]=i\hbar$, in the Heisenberg picture we have $\ddot{x}+\omega^{2}x=0$.  Decompose the position operator $x(t)$ in terms of the non-hermitian raising $(a^{\dagger})$ and lowering $(a)$ operators and mode function $f(t)$ as $x(t)=f(t)a+\bar{f}(t)a^{\dagger}$, where the over-bar represents complex conjugation, and the mode function satisfies the oscillator classical equation of motion $\ddot{f}(t)+\omega^{2}f(t)=0$.  Substituting into the commutation relation $\left[x,p\right]$ the above decomposition gives
\begin{equation}
\frac{m}{i\hbar}\left[x,\dot{x}\right]=\frac{m}{i\hbar}\left(f(t)\dot{\bar{f}}(t)-\bar{f}(t)\dot{f}(t)\right)\left[a,a^{\dag}\right]=1.
\end{equation}
Demanding the commutation relation $\left[a,a^{\dagger}\right]=1$ for all times, we have $\langle f,f\rangle=1$ and $\langle f,\bar{f}\rangle=0$, i.e. the mode functions $f(t)$ and $\bar{f}(t)$ are orthonormal in terms of the inner-product\footnote{In quantum field theory, the generalization of Eq.~(\ref{eq:kg}) to spacetimes where the dimensionality is larger than the zero-dimensional harmonic oscillator considered here is called the Klein-Gordon inner-product.}
\begin{equation}\label{eq:kg}
\langle f,g\rangle\equiv \frac{im}{\hbar}\left[\bar{f}(t)\dot{g}(t)-g(t)\dot{\bar{f}}(t)\right].
\end{equation} 
The ladder operators may then be defined in terms of this inner-product as $a=\langle f, x\rangle$ and $a^{\dagger}=-\langle \bar{f}, x\rangle$.  

Specifying the ground state of the system is equivalent to fixing the form of the mode function $f(t)$.  For the simple harmonic oscillator, the ground state can be defined with respect to the ladder operators as the state for which $a|0\rangle=0$.  Demanding this ground state be an eigenstate of the Hamiltonian $H|0\rangle=E|0\rangle$ gives the mode function equation of motion via
\begin{eqnarray}\label{eq:H}
&H|0\rangle&=\left(\frac{m\dot{x}^{2}}{2}+\frac{m\omega^{2}x^{2}}{2}\right)|0\rangle\\
&=&\frac{m}{2}\left\{\left[\dot{f}(t)a+\bar{\dot{f}}(t)a^{\dag}\right]^{2}+\omega^{2}\left[f(t)a+\bar{f}(t)a^{\dag}\right]^{2}\right\}|0\rangle \nonumber\\
&=&\frac{m}{\sqrt{2}}\overline{\left[\dot{f}(t)^{2}+\omega^{2}f(t)^{2}\right]}|2\rangle+\frac{m}{2}\left[\left|\dot{f}(t)\right|^{2}+\omega^{2}\left|f(t)\right|^{2}\right]|0\rangle. \nonumber
\end{eqnarray}
Since the term proportional to $|2\rangle$ must vanish, it follows that $\dot{f}(t)=\pm i \omega f(t)$ with normalization $\left|f(t)\right|^{2}=\hbar/(2m\omega)$ and inner-product $\langle f,f \rangle=\mp 1$.  Positivity of the inner-product selects the solution $f(t)=x_{\rm zp}\exp(-i\omega t)$ where $x_{\rm zp}=\sqrt{\hbar/2m\omega}$ is the zero-point uncertainty in the oscillator's position.  This is designated the ``positive frequency" solution\footnote{A complex function $f(t)=\frac{1}{\sqrt{2\pi}}\int_{-\infty}^{\infty}d\omega\, g(\omega)e^{-i\omega t}$ is said to be ``positive frequency" if it's Fourier transform $g(\omega)$ vanishes for all $\omega\le0$. In this case, $f(t)$ is composed solely of Fourier components of the form $e^{-i\omega t}$ where $\omega>0$.}, whereas $\bar{f}(t)=x_{\rm zp} \exp(+i\omega t)$ is the conjugate, ``negative frequency" solution.  Using Eq.~(\ref{eq:H}), it is straightforward to show that these mode functions lead to the canonical oscillator Hamiltonian $H=\hbar\omega\left(a^{\dag}a+1/2\right)$.  The position operator may then be written in the form
\begin{equation}\label{eq:x}
x(t)=x_{\rm zp}\left(e^{-i\omega t}a + e^{+i\omega t}a^{\dagger}\right),
\end{equation}
where we see that the positive (negative) frequency solution is associated with the annihilation (creation) operator.  

Now, suppose that the frequency of the harmonic oscillator is allowed to vary in time:
\begin{equation}\label{eq:time}
\ddot{x}+\omega(t)^{2}x=0, 
\end{equation}
such that the initial ``input" frequency is defined as $\omega(t\rightarrow -\infty)=\omega_{\rm in}$, and the final ``output" frequency is $\omega(t\rightarrow\infty)=\omega_{\rm out}$.  Here we assume that $\omega_{\rm out}$ differs from the input frequency $\omega_{\rm in}$.  These frequencies define two sets of ladder operators $a_{\rm in}$, $a_{\rm out}$, corresponding ground states $|0\rangle_{\rm in}$, $|0\rangle_{\rm out}$, and mode functions $f_{\rm in}(t)$, $f_{\rm out}(t)$, where from the above simple harmonic oscillator analysis, $\left.f_{\rm in}(t)\right|_{t\rightarrow -\infty} \sim \exp\left(-i\omega_{\rm in}t\right)$ and $\left.f_{\rm out}(t)\right|_{t\rightarrow +\infty}\sim \exp\left(-i\omega_{\rm out}t\right)$, with 
\begin{equation}
x(t)=f_{\mathrm{in}}(t)a_{\mathrm{in}}+\bar{f}_{\mathrm{in}}(t)a_{\mathrm{in}}^{\dagger}=f_{\mathrm{out}}(t)a_{\mathrm{out}}+\bar{f}_{\mathrm{out}}(t)a_{\mathrm{out}}^{\dagger}.
\end{equation}
As a second-order differential equation, Eq.~(\ref{eq:time}) requires two linearly independent solutions to characterize the dynamics. Given that $f_{\rm in}$ is a solution to the oscillator equation and  $\langle f_{\mathrm{in}},\bar{f}_{\mathrm{in}}\rangle=0$, we may write the output state modes as a linear combination of the input state solutions, $f_{\rm out}=\alpha f_{\rm in}+\beta \bar{f}_{\rm in}$.  Substituting into Eq.~(\ref{eq:kg}), the coefficients are connected through the symplectic relation
\begin{equation}\label{eq:constraint}
\left|\alpha\right|^{2}-\left|\beta\right|^{2}=1.
\end{equation}
With $f_{\rm out}(t)$ expressed using input modes, the output state lowering operator  $a_{\rm out}=\langle f_{\rm out},x\rangle$ is then given as
\begin{equation}\label{eq:bogoliubov}
a_{\rm out}=\alpha a_{\rm in}-\bar{\beta}a_{\rm in}^{\dagger}.
\end{equation}
Assuming the oscillator is initially in the ground state $|0\rangle_{\rm in}$, the particle number expectation value at the output is $N_{\rm out}=\langle0|a^{\dagger}_{\rm out}a_{\rm out}|0\rangle_{\rm in}=\left|\beta\right|^{2}$.  Other than adiabatic changes from $\omega_{\rm in}$ to $\omega_{\rm out}$, $\beta$ is non-vanishing, and there is a finite probability of the oscillator being found in an excited state at the output; the average excitation number $N_{\rm out}$ is determined by the coefficient of the negative frequency ($a_{\rm in}^{\dag}$) coefficient in Eq.~(\ref{eq:bogoliubov}).

Equation~(\ref{eq:bogoliubov}) is an example of a larger class of transformation called Bogoliubov transformations, where the ladder operators in the output state may be written as a linear combination of \textit{both} initial state creation and annihilation operators with coefficients satisfying the constraint given in Eq.~(\ref{eq:constraint}).  All quantum amplification processes can be cast as Bogoliubov transformations \cite{leonhardt:2010}.  They therefore represent a useful generalized framework within which one may compare the various amplification methods. 

\section{Vacuum amplification}\label{sec:vacuum}
\begin{figure}[t]\begin{center}
\includegraphics[width=8.0cm]{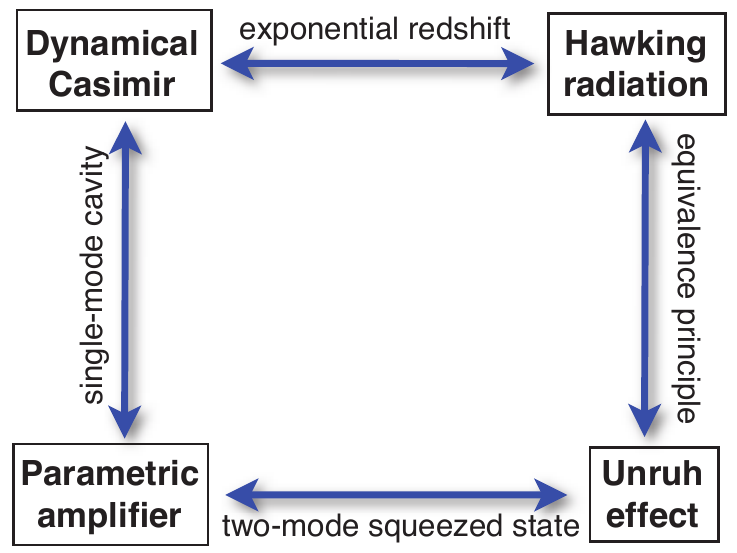}
\caption{(Color online) Relationships between quantum amplification mechanisms. 
Counterclockwise from the parametric amplifier: For a single mode of the Minkowski vacuum, the
non-degenerate parametric amplifier (NDPA) and Unruh effect (UE) share the same
form of Bogoliubov transformations resulting in both exhibiting a two-mode
squeezed state.  The UE is in turn connected to Hawking radiation (HR) through
the equivalence principle relating inertial and gravitational acceleration.  The
exponential red-shifting (Doppler shift) of the field modes near the black hole
horizon results in Bogoliubov transformations that are identical to those for
the dynamical Casimir effect (DCE), provided the mirror's trajectory is given by
Eq.~(\ref{eq:receding-mirror}). Here, one obtains an identical Doppler shift,
leading to a thermal spectrum for the emitted radiation.  Finally, the DCE and a
degenerate parametric amplifier (DPA) can be related by considering the case of
a single-mode cavity with a sinusoidally time-dependent boundary condition.}
\label{fig:relations}
\end{center}
\end{figure}
In this section we review the main mechanisms by which vacuum fluctuations are amplified into photons: the parametric amplifier (PA), Unruh effect (UE), Hawking radiation (HR), and the dynamical Casimir effect (DCE).  Although these effects were first discovered in seemingly unrelated contexts, the universal description of quantum amplification provided by Bogoliubov transformations suggests these mechanisms are in fact closely related.  Before exploring these effects in detail, we wish to draw the reader's attention to Fig.~(\ref{fig:relations}), which highlights in summary form the key conditions under which the various amplification mechanisms may be related. Fig.~(\ref{fig:relations}) serves to motivate the subsequent sections, where the depicted relationships are made explicit, and thus linked back to the parametric amplifier, our main objective.

\subsection{Parametric amplification}\label{sec:paramp}
\begin{figure}[t]
\includegraphics[width=6.0cm]{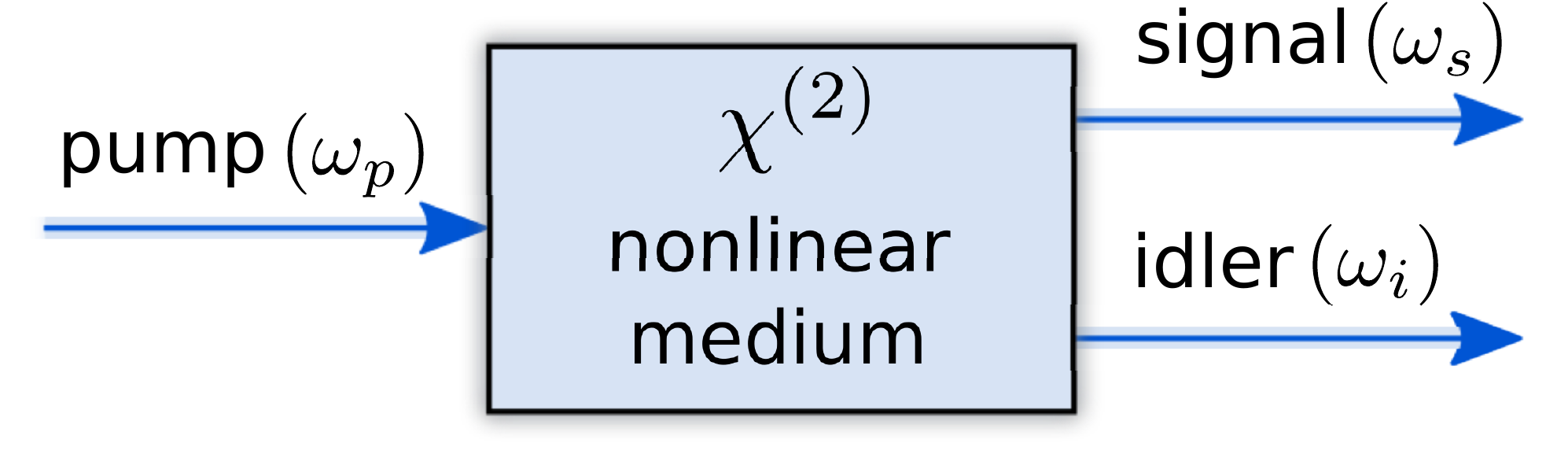}
\caption{(Color online) The principle of a parametric amplifier: a pump photon is down-converted by a nonlinear medium into a signal and
an idler photon, whose frequencies add up to that of the pump photon.}
\label{fig:parametric-amplification-v1}
\end{figure}

All quantum amplifiers are inherently nonlinear systems \cite{clerk:2010}.  One of the simplest nonlinear interactions, indicated in Fig.~\ref{fig:parametric-amplification-v1}, involves a pump photon of frequency $\omega_{p}$ being converted into two photons denoted the signal ($\omega_{s}$) and idler ($\omega_{i}$), obeying the frequency relation $\omega_{p}=\omega_{s}+\omega_{i}$.  This process is known as parametric down conversion and occurs in a dielectric medium with a $\chi^{(2)}$ nonlinearity, the first nonlinear susceptibility in a medium without inversion symmetry \cite{boyd:2008}.

When a cavity is driven by a classical pump such as a laser or microwave generator that is not significantly attenuated by the loss of photons via the down-conversion process, this nonlinear interaction can be described by an effective Hamiltonian which, in the rotating frame, takes the form
\begin{equation} \label{eq:pa-hamiltonian}
H= i\hbar\eta(b_{\rm s}^\dag b_{\rm i}^\dag -b_{\rm s} b_{\rm i}),
\end{equation}
where $\eta$ is the pump amplitude dependent coupling strength, and the subscripts denote signal ($s$) and idler ($i$) modes respectively.  In the special case that the signal and idler modes coincide $b_{s}=b_{i}=b$, Eq.~(\ref{eq:pa-hamiltonian}) describes a degenerate parametric amplifier (DPA) 
where the pump drives the cavity mode at twice it's resonance frequency.  The Heisenberg equations of motion that follow from the Hamiltonian Eq.~(\ref{eq:pa-hamiltonian}) lead to the time-evolution of the cavity mode operator
\begin{equation}\label{eq:dpa}
b(t)=b(0)\cosh\left(2\eta t\right)+b(0)^{\dag}\sinh\left(2\eta t\right),
\end{equation}
which is characteristic of a squeezing transformation \cite{walls:2008}.  Comparison with Eq.~(\ref{eq:bogoliubov}) indicates that Eq.~(\ref{eq:dpa}) is in fact a Bogoliubov transformation with $\alpha =\cosh\left(2\eta t\right)$ and $\beta=\sinh\left(2\eta t\right)$.  These coefficients are easily seen to satisfy the symplectic relation Eq.~(\ref{eq:constraint}).   Assuming the mode is initially in the ground state, the number of excitations at later times is calculated from the coefficient of the negative frequency component ($b^{\dag}$) to be $N=\left<b^{\dag}(t)b(t)\right>=\left|\beta\right|^{2}=\sinh^{2}(2\eta t)$.  The fact that $N$ grows as a function of time, even when starting from the vacuum state, is a purely quantum mechanical manifestation of parametric amplification of vacuum fluctuations.  The effects of the squeezing transformation can be seen by defining quadrature amplitudes $X_{1}=b+b^{\dag}$ and $X_{2}=(b-b^{\dag})/i$ related to the mode's position and momentum operators respectively.  By analogy with the classical parametric amplifier in Eq.~(\ref{eq:para-swing}), the DPA is a phase-sensitive amplifier, amplifying one quadrature of motion $X_{1}(t)=e^{2\eta t}X_{1}(0)$, while attenuating the other quadrature $X_{2}(t)=e^{-2\eta t}X_{2}(0)$.

The more general case of independent signal and idler modes represents a phase-sensitive amplification process know as the non-degenerate parametric amplifier (NDPA).  The time-evolution of the signal and idler modes under the influence of the Hamiltonian (\ref{eq:pa-hamiltonian}) is described by a pair of Bogoliubov transformations
\begin{eqnarray}\label{eq:ndpa}
b_{s}(t)&=&b_{s}(0)\cosh\left(\eta t\right)+b^{\dagger}_{i}(0)\sinh\left(\eta t\right)     \nonumber \\ 
b_{i}(t)&=&b_{i}(0)\cosh\left(\eta t\right)+b^{\dagger}_{s}(0)\sinh\left(\eta t\right),
\end{eqnarray}
where again, the number of quanta in each of the modes is easily calculated  from the coefficients of the creation operator components, $N_{s}= N_{i} = \sinh^2(\eta t)$, assuming both modes are initially in their ground states.

In the Schr\"odinger picture, the wave function for the signal and idler modes is
\begin{eqnarray}
\label{eq:par-amp-wave-function}
\left|\Psi(t)\right> = \frac{1}{\cosh\eta t}\sum_{n=0}^{\infty} \left(\tanh\eta t\right)^n \left|n\right>_{s}\otimes\left|n\right>_{i},
\end{eqnarray}
where $\left|n\right>_{s}\otimes \left|n\right>_{i}$ corresponds to $n$ photons in each of the signal and idler modes.  Given the form of the transformation in Eq.~(\ref{eq:ndpa}), the resulting state of the system (\ref{eq:par-amp-wave-function}) is a two-mode squeezed state, where $\eta t$ plays the role of squeezing parameter.  In contrast to the DPA, the squeezing of the NDPA does not occur in a single mode, but rather in the composite system formed by the combined signal and idler modes \cite{walls:2008}.  The two-mode squeezed state (\ref{eq:par-amp-wave-function}) is an example of an Einstein-Podolsky-Rosen (EPR) state \cite{einstein:1935} where the correlations between the signal and idler modes is stronger than that allowed by classical theory \cite{reid:1988}.

In cases where, either by choice or design, only one of the two modes is accessible, measurements on the remaining mode do not contain enough information to reconstruct Eq.~(\ref{eq:par-amp-wave-function}).  Given the close relationship between information and entropy, this loss of information is encoded in the entropic properties of the measured single-mode state.  As a bipartite system, the entropy of the measured mode may be calculated via the von Neumann entropy $S$ of the reduced density matrix obtained by tracing over the unobserved mode, also referred to as the entanglement entropy \cite{nielson:2000}.  With the signal $(s)$ mode as the observed mode, tracing over the unobserved idler $(i)$ mode, we obtain for the entanglement entropy $S=-\mathrm{Tr}\rho_{s}\ln\rho_{s}$:
\begin{equation}\label{eq:thermal-osc}
S=-\ln\left[1-e^{-\hbar\omega_{s}/k_{\mathrm{B}}T(t)}\right]-\frac{\hbar\omega_{s}}{k_{\mathrm{B}}T(t)}\left[1-e^{\hbar\omega_{s}/k_{\mathrm{B}}T(t)}\right]^{-1},
\end{equation}
which is just the thermal entropy (neglecting the overall Boltzmann factor) of a quantum harmonic oscillator with temperature $T(t)$ related to the squeezing parameter via
\begin{equation}\label{eq:tt}
 \tanh^{2} \eta t = \exp\left[-\frac{\hbar\omega_{s}}{k_{\mathrm{B}}T(t)}\right].  
\end{equation}
Therefore, the non-vanishing entropy or equivalently information lost, by  tracing over one of the two modes in a particle pair squeezed state (\ref{eq:par-amp-wave-function}) signals that  the remaining mode is in a mixed, thermal state \cite{barnett:1985,yurke:1987}.  

To understand the origin of the thermal state (\ref{eq:thermal-osc}) we note that, as unbounded harmonic oscillator mode systems, both the signal and idler states contain an infinite ladder of energy levels.  In order to obtain a finite value for the entropy, the average energy, or equivalently number of particles, in each of the modes must also be specified \cite{barnett:1989,barnett:1991}. Although we do not know the quantum state of the idler mode after tracing over it in Eq.~(\ref{eq:par-amp-wave-function}), the correlations between photon number in the signal and idler modes, enforced by energy conservation, gives us implicit knowledge about the average energy of the idler state.  Knowing only the energy of the idler mode, maximizing the entropy, or equivalently minimizing the information, of the idler state with respect to this constraint yields the thermal state entropy of Eq.~(\ref{eq:thermal-osc}).  The bipartite structure of Eq.~(\ref{eq:par-amp-wave-function}) demands that this same value of the entropy hold for the measured signal mode as well.

\subsection{The Unruh effect}\label{sec:unruh}
Conceptually, perhaps the simplest way to generate particles from the vacuum is for an observer to accelerate.  Unlike an inertial observer in Minkowski space, an observer undergoing constant acceleration is out of causal contact with a portion of the entire space-time due to the presence of a horizon.  As a result, the initially pure Minkowski quantum vacuum state will appear to the observer to be in a mixed thermal state \cite{unruh:1976,crispino:2008}.

Before exploring this Unruh effect (UE) \cite{unruh:1976}, we need to define what is meant by ``an observer".  As the name suggests, an observer should be a witness to the dynamics under consideration.  As our focus here is on the generation of particles from the quantum vacuum, the observer is ideally represented by a particle-detector.  Although a variety of model systems may be used for the particle-detector, for our purposes the observer will be represented as a two-level system, or qubit, detector with ground $|0\rangle$ and first-excited $|1\rangle$ energy levels separated by an energy $\hbar\omega_{01}$.  In addition, we will assume a point-like detector that is linearly-coupled to the operators representing the quantized field or cavity mode of interest \cite{birrell:1982}.  We will further suppose that the detector is weakly coupled to the field modes so as to allow the transition probabilities between the qubit ground and excited states to be calculated perturbatively \cite{clerk:2010}.  Our choice of two-level detector will be further motivated in Sec.~\ref{sec:sc-circuits}, where we discuss the use of a superconducting phase-qubit as a single-shot microwave photon counter \cite{chen:2010}.

\begin{figure}[t]\begin{center}
\includegraphics[width=8.0cm]{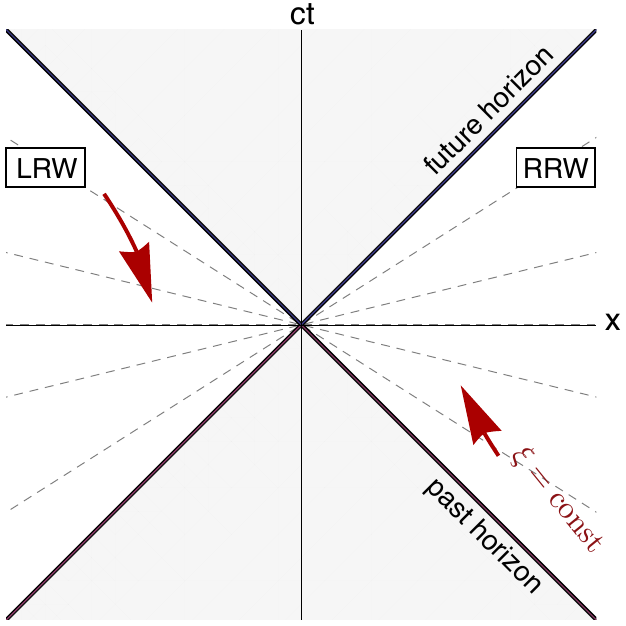}
\caption{(Color online) Paths of accelerated observers in Rindler coordinates $(c \tau,\xi)$ with proper time $\tau$ and constant acceleration $a=c^{2}/\xi$ as viewed in Minkowski space-time with coordinates $(ct,x)$.  Lines (dashed) of constant proper time $\tau$ are also indicated. Observers in the right Rindler wedge (RRW) are out of causal contact with the left Rindler wedge (LRW) due to the presence of a horizon at $ct=\pm x$.  Arrows give the direction of increasing proper time in each Rindler wedge.}
\label{fig:rindler}
\end{center}
\end{figure}

Having established the definition of an observer, let us now consider the worldline of an observer undergoing a constant proper acceleration $a$.  In Minkowski coordinates $(ct,x)$, the paths of observers with constant acceleration are hyperbolas in space-time as seen in Fig.~\ref{fig:rindler}.  For $a>0$, these paths trace out a section of Minkowski space known as the Right Rindler wedge (RRW) defined by the relation $|ct|<x$, and may be described using the Rindler coordinates, $(c\tau,\xi)$, describing the observer's path through Minkowski spacetime as viewed by the observer herself, and defined through the relations
\begin{equation}\label{eq:rindler-eqs}
ct=\xi\sinh\left(\frac{a\tau}{c}\right)\   \ ;\   \ x=\xi\cosh\left(\frac{a\tau}{c}\right),
\end{equation}
where $\tau$ is the observer's proper time and $\xi=c^{2}/a$ is the distance from the vertex (i.e. the closest point to the origin) of the observer's motion to the origin. 
In switching to Rindler coordinates, the observer moves only in the direction of increasing proper time $\tau$, while the spatial coordinate $\xi$ remains constant, thus greatly simplifying the resulting equations of motion.  Rewriting the Minkowski metric $ds^{2}=-c^{2}dt^{2}+dx^{2}$ in Rindler coordinates gives the Rindler metric 
\begin{equation}\label{eq:rindler}
ds^{2}=-\left(\alpha\xi\right)^{2}d\tau^{2}+d\xi^{2},
\end{equation}
where $\alpha=a/c$ is a parameter characterizing the proper acceleration.  Relative to the RRW, we may also define mathematically a second Left Rindler wedge (LRW) with $x<|ct|$ by reflecting the RRW across the $ct$-axis ($t\rightarrow -t$) and then across the x-axis ($x\rightarrow-x$) \cite{birrell:1982}.  This change in sign for the time-coordinate in the LRW causes the proper-time $\tau$ to run backwards in Minkowski time $t$ as shown in Fig.~\ref{fig:rindler}. The two Rindler wedges are causally disconnected from each other as a result of a horizon located on the lightcone $ct=\pm x$.  Trajectories of observers are asymptotically bound by this lightcone for $\tau\rightarrow-\infty$ and $\tau\rightarrow\infty$ where the observer's velocity approaches the speed of light. These limits represent the past and future horizons, respectively.  Likewise, the path of an observer undergoing infinite acceleration $a\rightarrow\infty$ ($\xi\rightarrow 0$) lies on the horizon of the RRW, as may be checked from Eq.~(\ref{eq:rindler-eqs}).

In order to describe the Minkowski vacuum as seen by the accelerating observer, we will proceed in a manner similar to the time-dependent oscillator example in Sec.~\ref{sec:amp-intro}.  First, we find the mode functions and their associated vacuum states for a scalar quantum field in both the Minkowski and Rindler spacetimes.  We then calculate the Bogoliubov transformations linking the Minkowski and Rindler creation and annihilation operators.  With the Bogoliubov transformations in hand, the quantum state seen by a RRW observer is readily obtained.

Analogously to the position operator for the harmonic oscillator in Eq.~(\ref{eq:x}), a scalar field in Minkowski spacetime may be expanded as a infinite sum of positive and negative frequency components,
\begin{equation}\label{eq:sum}
\phi=\sum_{j}u^{\mathrm{M}}_{\omega_{j}}a^{\mathrm{M}}_{\omega_{j}}+\bar{u}^{\mathrm{M}}_{\omega_{j}}a^{\mathrm{M},\dag}_{\omega_{j}},
\end{equation}
where the positive-frequency, orthonormal mode field functions are solutions to the 2D Minkowski wave equation
\begin{equation}\label{eq:wave-equation}
\left[\frac{1}{c^{2}}\frac{\partial^{2}}{\partial t^{2}}-\frac{\partial^{2}}{\partial x^{2}}\right]\phi=0,
\end{equation}
and given by the plane-waves
\begin{equation}\label{eq:plane}
u_{\omega_{j}}^{\mathrm{M}}=\frac{1}{\sqrt{4\pi\omega_{j}}}e^{ik_{j}x-i\omega_{j}t},
\end{equation}
with $\omega_{j}=c|k_{j}|$ and $-\infty \le j \le \infty$, where the superscript $\mathrm{M}$ signifies belonging to the Minkowski spacetime.  The Minkowski vacuum state $|0\rangle^{\mathrm{M}}=\prod_{j}|0_{\omega_{j}}\rangle^{\mathrm{M}}$ is defined with respect to the positive frequency modes as the state that is annihilated by all lowering operators $a^{\mathrm{M}}_{\omega_{j}}$, i.e.  $a^{\mathrm{M}}_{\omega_{j}}|0\rangle^{\mathrm{M}}=0$ for all $j$.

Of course, the accelerated observer may also define a vacuum state for the quantum field in the Rindler spacetime using the associated Rindler coordinates.  Here, the orthonormal mode functions may be found by solving the 2D wave equation Eq.~(\ref{eq:wave-equation}) expressed in Rindler coordinates via Eq.~(\ref{eq:rindler-eqs}).  As a static spacetime, the Rindler metric (\ref{eq:rindler}) admits a natural vacuum state $|0\rangle^{\mathrm{R}}=\prod_{j}|0_{\omega_{j}}\rangle^{\mathrm{R}}$ in the RRW with respect to the positive frequency Rindler modes $u_{\omega_{j}}^{\mathrm{R}}\propto \exp\left(-i\omega_{j}\tau\right)$.  Note that the notion of positivity for the Rindler modes is with respect to the observer's proper time $\tau$.  The Rindler coordinates ($c\tau,\xi$) in the LRW are completely independent of those in the RRW, giving rise to independent vacuum states for the LRW and RRW spacetimes.  Again, the LRW vacuum state $|0\rangle^{\mathrm{L}}=\prod_{j}|0_{\omega_{j}}\rangle^{\mathrm{L}}$ is defined with respect to positive-frequency Rindler modes $u_{\omega_{j}}^{\mathrm{L}}$.  However, as a consequence of the reflection $t\rightarrow -t$ used in defining the LRW, the notion of positive and negative frequencies is switched in the LRW. The result is a vacuum state in the LRW that is defined with respect to positive frequency modes $u_{\omega_{j}}^{\mathrm{L}}\propto \exp\left(i\omega_{j}\tau\right)$.

The Rindler modes $u_{\omega_{j}}^{\mathrm{R}}$,$u_{\omega_{j}}^{\mathrm{L}}$ and Minkowski modes $u_{\omega_{j}}^{\mathrm{M}}$ are not independent.  Rather, they represent different expansions of the scalar field $\phi$ and therefore are related by a change of basis.  As seen in Fig.~\ref{fig:rindler}, the RRW (or LRW) covers only $1/4$ of the entire Minkowski spacetime and as a result the Rindler modes in this region are not enough to reconstruct the entire Minkowski spacetime modes \cite{unruh:1976,birrell:1982}.  We can however take a linear combination of modes from both Rindler wedges and, through analytic continuation \cite{boulware:1975}, cover the entire spacetime.  In taking linear combinations of modes from the LRW and RRW, we have effectively mixed positive and negative frequency components.  Given our discussion on Bogoliubov transformations in Sec.~\ref{sec:amp-intro}, when expressed in this combined Rindler basis, one should expect the Minkowski vacuum viewed by the accelerating observer to contain particles.  As we shall see, this is indeed the case.

The general expansion of the Minkowski modes in Rindler modes reads,
\begin{equation}\label{eq:mode-relations}
u_{\omega_{j}}^{\mathrm{M}}= \sum_{i}\alpha_{ij}^{\mathrm{R}}u_{\omega_{i}}^{\mathrm{R}}+\bar{\beta}_{ij}^{\mathrm{R}}\bar{u}_{\omega_{i}}^{\mathrm{R}}+\alpha_{ij}^{\mathrm{L}}u_{\omega_{i}}^{\mathrm{L}}+\bar{\beta}_{ij}^{\mathrm{L}}\bar{u}_{\omega_{i}}^{\mathrm{L}}
\end{equation}
where $\alpha_{ij}^{\mathrm{R,L}}$ and $\beta_{ij}^{\mathrm{R,L}}$ are Bogoliubov transformation matrices  with coefficients given by the Klein-Gordon inner-product between Minkowski and Rindler modes
\begin{equation}\label{eq:product}
\alpha_{ij}^{\mathrm{R,L}}=\left<u_{\omega_{i}}^{\mathrm{R,L}},u_{\omega_{j}}^{\mathrm{M}}\right>;\  \ \beta_{ij}^{\mathrm{R,L}}=-\left<u_{\omega_{i}}^{\mathrm{R,L}},\bar{u}_{\omega_{j}}^{\mathrm{M}}\right>.
 \end{equation}
The connection between ladder operators and mode functions allows us to use Eq.~(\ref{eq:mode-relations}) to establish the Bogoliubov transformation between Minkowski and Rindler ladder operators as
\begin{equation}\label{eq:oper-relations}
a_{\omega_{j}}^{\mathrm{M}}= \sum_{i}\alpha_{ij}^{\mathrm{R}}a_{\omega_{i}}^{\mathrm{R}}+\bar{\beta}_{ij}^{\mathrm{R}}a_{\omega_{i}}^{\dag,\mathrm{R}}+\alpha_{ij}^{\mathrm{L}}a_{\omega_{i}}^{\mathrm{L}}+\bar{\beta}_{ij}^{\mathrm{L}}a_{\omega_{i}}^{\dag,\mathrm{L}}.
\end{equation}

Although we can explicitly evaluate Eq.~(\ref{eq:product}) to obtain the Bogoliubov transformation matrices in (\ref{eq:oper-relations}), the result does not elucidate the underlying physics of the amplification process as a single Minkowski mode $\omega_{j}$ will transform into a continuum of Rindler modes.  Instead, we note that the Minkowski vacuum state $|0\rangle^{\mathrm{M}}$ is defined with respect to the positive frequency modes, $u_{\omega_{j}}^{\mathrm{M}}$, and any other set of basis mode functions constructed from a linear combination of these Minkowski modes will leave the vacuum state $|0\rangle^{\mathrm{M}}$ unchanged \cite{birrell:1982}.  We therefore construct the Unruh basis \cite{unruh:1976} set of mode functions $\left\{v_{\omega_{j}}^{(1),\mathrm{M}},v_{\omega_{j}}^{(2),\mathrm{M}}\right\}$, from linear combinations of positive frequency Minkowski modes
\begin{equation}
v_{\omega_{j}}^{(1),\mathrm{M}}=\sum_{i}\epsilon^{(1)}_{ij}u_{\omega_{i}}^{\mathrm{M}};\  \ v_{\omega_{j}}^{(2), \mathrm{M}}=\sum_{i}\epsilon^{(2)}_{ij}u_{\omega_{i}}^{\mathrm{M}}
\end{equation}
such that, when expanded in the Rindler modes $\left\{u_{\omega_{j}}^{\mathrm{R}},u_{\omega_{j}}^{\mathrm{L}}\right\}$, diagonalizes the Bogoliubov transformation matrices $\alpha_{ij}$ in Eq.~(\ref{eq:oper-relations}).  For the annihilation operators $b_{\omega_{j}}^{(1),\mathrm{M}},b_{\omega_{j}}^{(2),\mathrm{M}}$ associated with mode functions $v_{\omega_{j}}^{(1),\mathrm{M}},v_{\omega_{j}}^{(2),\mathrm{M}}$, this procedure yields the Bogoliubov transformations for the Rindler operators \cite{unruh:1976,birrell:1982}
\begin{eqnarray}\label{eq:unruh-bogo}
b_{\omega_{j}}^{(1),\mathrm{M}}&=&a^{\mathrm{R}}_{\omega_{j}}\cosh\left(r\right)+a^{\dag,\mathrm{L}}_{\omega_{j}}\sinh\left(r\right)     \nonumber \\ 
b_{\omega_{j}}^{(2),\mathrm{M}}&=&a^{\mathrm{L}}_{\omega_{j}}\cosh\left(r\right)+a^{\dag,\mathrm{R}}_{\omega_{j}}\sinh\left(r\right),
\end{eqnarray}
with the effective squeezing parameter $r$ defined by $\tanh r =\exp\left(-\pi\omega_{j}/\alpha\right)$.  In the Unruh basis we have a monochromatic Bogoliubov transformation relating a single Minkowski mode $\omega_{j}$ to the same mode in both the left and right Rindler wedges.  More importantly, the Bogoliubov transformations (\ref{eq:unruh-bogo}) are of the same form as the transformations for the NDPA in Eq.~(\ref{eq:ndpa}).  Thus we establish the connection between the NDPA and the UE summarized in Fig.~(\ref{fig:relations}). 

For a single mode of the Minkowski vacuum $|0_{\omega_{j}}\rangle^{\mathrm{M}}$, the Bogoliubov transformations in Eq.~(\ref{eq:unruh-bogo}) lead to the two-mode squeezed  state for the Rindler modes
\begin{equation}\label{eq:unruh-state}
|0_{\omega_{j}}\rangle^{\mathrm{M}}= \frac{1}{\cosh r}\sum_{n=0}^{\infty} \left(\tanh r \right)^n | n_{\omega_{j}}\rangle^{\mathrm{L}}\otimes | n_{\omega_{j}}\rangle^{\mathrm{R}}.
\end{equation}
From the viewpoint of the observer in the RRW, the presence of the horizon prevents access to the modes in the LRW and they must be traced over in Eq.~(\ref{eq:unruh-state}).  By analogy with the NDPA in Sec.~\ref{sec:paramp}, the observed mode in the RRW are in a thermal state with temperature related to the squeezing parameter $r$ as follows: 
\begin{equation}
\tanh^{2}\left(r\right)=e^{-2\pi\omega/\alpha}=\exp\left(-\frac{\hbar\omega}{k_{\mathrm{B}}T_{\mathrm{U}}}\right), 
\end{equation}
where the Unruh temperature is
\begin{equation}\label{eq:tu}
T_{\mathrm{U}}=\frac{\hbar\alpha}{2\pi k_{\mathrm{B}}},  
\end{equation}
in terms of the proper acceleration parameter $\alpha=a/c$.  Here, the energy required to generate particles from the vacuum comes from the work needed to maintain the observers constant acceleration.  Like the parametric amplifier, Sec.~\ref{sec:paramp}, we have implicitly assumed the energy of the accelerating observer is unaffected by the creation of particles.  The transfer of energy to the field modes is quite natural given that our detector is linearly-coupled to the operators representing the quantized scalar field.  As discussed earlier, these field modes are not local to the observer, but rather form a basis set covering the entire spacetime.  As a result, the full spacetime of a Rindler observer is in a thermal state characterized by the Unruh temperature Eq.~(\ref{eq:tu}).

An equivalent way to understand the origin of the Unruh temperature $T_{\mathrm{U}}$ is to consider the effect of the horizon in the accelerating reference frame on a monochromatic plane wave with frequency $\Omega$ moving in the $x$-direction of Minkowski space, $\phi(x,t)=\exp\left[-i\Omega\left(t-x/c\right)\right]$.  From the viewpoint of the accelerating observer, this wave may be expressed via Eq.~(\ref{eq:rindler-eqs}) as 
\begin{eqnarray}\label{eq:rindler-mode}
\phi(\tau)&=&\exp\left\{\frac{-i\Omega\xi}{c}\left[\sinh\left(\alpha\tau\right)-\cosh\left(\alpha\tau\right)\right]\right\}\nonumber \\
&=&\exp\left[i\frac{\Omega}{\alpha}\left(e^{-\alpha\tau}\right)\right],
\end{eqnarray}
where we have used $\xi=c^{2}/a$.  We see that the wave is no longer monochromatic, but is rather exponentially red-shifted (Doppler shifted) with an e-folding time determined by the observer's acceleration $\alpha$.  Upon Fourier transforming Eq.~(\ref{eq:rindler-mode}), $f(\omega)=\frac{1}{\sqrt{2\pi}}\int_{-\infty}^{\infty}d\tau\, \phi(\tau)e^{+i\omega \tau}$, the effect of this red-shift can be seen in the resulting power spectrum, $P\left(\omega\right)=\left|f\left(\omega\right)\right|^{2}$, which does not vanish at negative frequencies: 
\begin{equation}
P(-\omega)=\left|f(-\omega)\right|^{2}=\frac{2\pi}{\omega\alpha}\frac{1}{e^{\frac{2\pi\omega}{\alpha}}-1};\ \ \omega>0.
\end{equation}
Comparing with a Planck distribution, we again recover the Unruh temperature Eq.~(\ref{eq:tu}) \cite{padmanabhan:2005}.  

For the two-level observer/detector, the ratio of the power spectrum $P(\omega)$ evaluated at negative and positive qubit transition frequencies, $\mp\omega_{01}$ respectively, can be related to the Fermi golden rule transition rates $\Gamma$ between ground and excited-state energy levels \cite{clerk:2010}:
\begin{equation}\label{eq:detailed}
\frac{P(-\omega_{01})}{P(\omega_{01})}=\frac{\Gamma_{|0\rangle\rightarrow |1\rangle}}{\Gamma_{|1\rangle\rightarrow |0\rangle}}=\exp\left[\frac{-\hbar \omega_{01}}{k_{\mathrm{B}}T_{\mathrm{U}}}\right],
\end{equation}
which is identical to the detailed balance relation for transition rates in a thermal environment.  In this way, the negative frequency terms represent the absorption of energy by the observer from the environment, whereas positive frequencies indicate emission.  The excitation of the two-level detector can only occur if there are particles in the field mode to which it is coupled.  The negative-frequency components signal the presence of particles as seen by the observer, and the departure from the Minkowski vacuum state.  From the viewpoint of the accelerated observer, Eq.~(\ref{eq:detailed}) indicates that there is no difference between the transformed Minkowski vacuum state and a thermal environment at the Unruh temperature.  We must therefore consider the Unruh temperature as corresponding to the actual physical temperature of the environment as seen by the observer. 

Although the UE shares many features with the NDPA in Sec.~\ref{sec:paramp}, there are several important differences.  For a constant acceleration, the squeezing parameter $r$, and therefore Unruh temperature $T_{\mathrm{U}}$, is time independent.  Likewise, Eq.~(\ref{eq:unruh-state}) shows that $T_{\mathrm{U}}$ is the same for any choice of mode frequency $\omega_{j}$.  This is in contrast to the parametric amplifier where the effective temperature is time dependent [Eq.~(\ref{eq:tt})] due to particle build up and with rates that depend on the mode coupling strength, pump amplitude and frequency \cite{leonhardt:2010}.  Furthermore, in contrast to the NDPA where in principle both modes of the two-mode squeezed state (\ref{eq:par-amp-wave-function}) can be measured, the existence of a horizon for the accelerating observer allows only those modes in the RRW to be measured.  The resulting thermal environment is of fundamental importance to quantum information and entanglement in relativistic systems \cite{hartle:1995,alsing:2003,fuentes:2005,peres:2004}.

\subsection{Hawking radiation}\label{sec:hawking}
\begin{figure}[t]\begin{center}
\includegraphics[width=8.0cm]{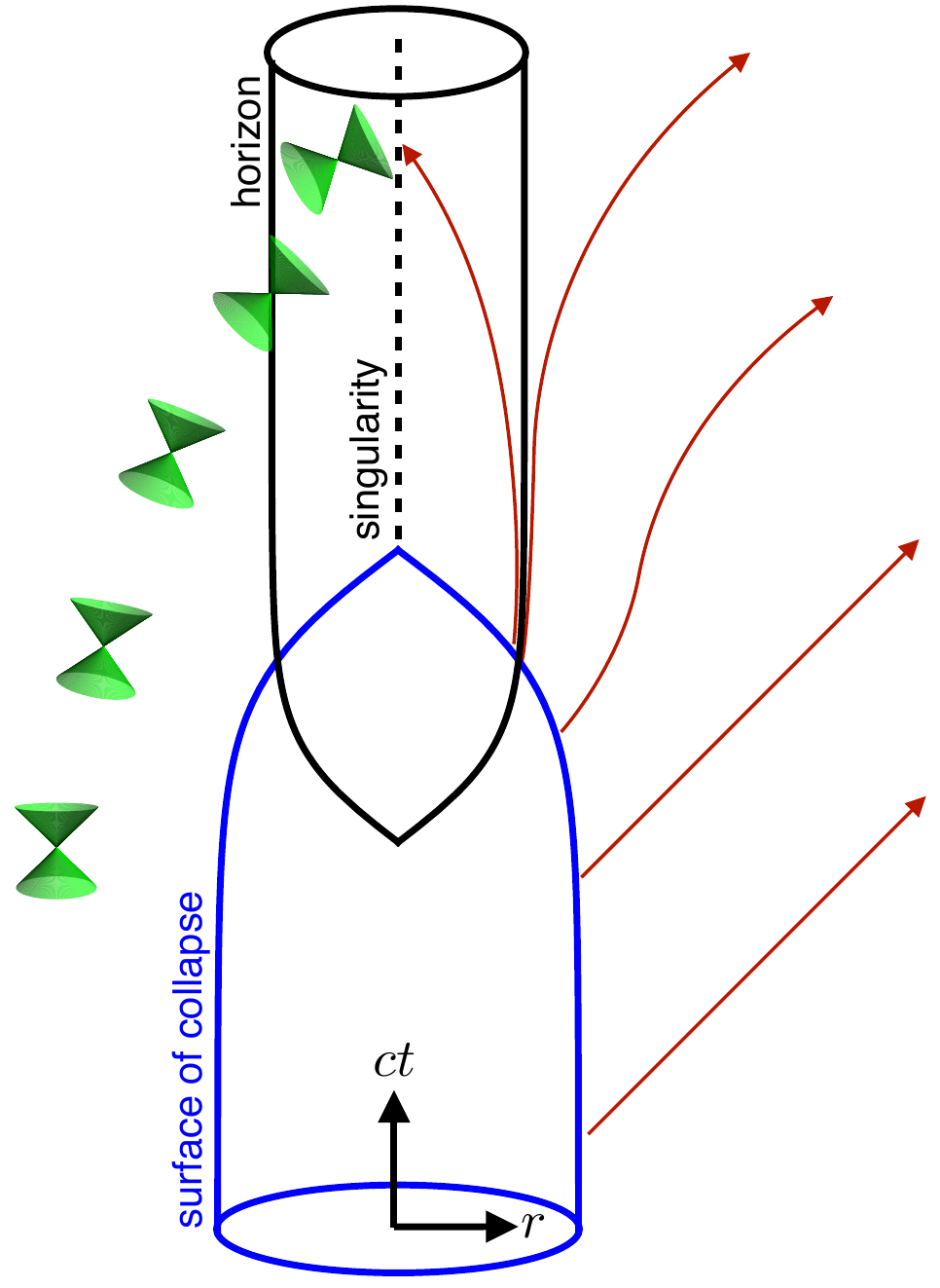}
\caption{(Color online) Formation of a horizon (black) by the gravitational collapse of a spherical object.  Before the horizon forms, light rays (red) leaving the surface of the object (blue) are free to propagate out to spatial infinity.  In contrast, once the mass of the body is within the Schwarzschild radius $r_{s}=2GM/c^{2}$, light rays are trapped behind the horizon and eventually encounter the singularity (dashed line).  The horizon demarcates the last light ray able to escape from the surface to infinity and the first trapped ray inside the radius $r_{s}$.  Equivalently, the horizon can be characterized by looking at the causal structure of spacetime indicated by light-cones (green) that give the direction of propagation for light rays at a given point.  As one approaches the horizon, the light-cone begins to tilt toward the black hole singularity.  On the horizon, the light-cone aligns along the $ct$-direction such that a light ray emitted from the horizon is stationary in space.  As the time-component of the metric vanishes on the horizon, a light ray on the horizon also appears frozen in time.  Inside the horizon, even time itself is points toward the singularity, so that nothing can escape.}
\label{fig:collapse}
\end{center}
\end{figure}

One of the most astonishing predictions of general relativity is that of a black hole, a region of spacetime where gravity is so strong that not even light can escape its pull.  When viewed by an observer at rest far from the black hole, a non-rotating, uncharged black hole with mass $M$ can be described by the Schwarzschild metric  
\begin{equation}\label{eq:Schwarzschild}
ds^{2}=-\left(1- \frac{r_{s}}{r}\right)c^{2}dt^{2}+\left(1-\frac{r_{s}}{r}\right)^{-1}dr^{2}+r^{2}d\Omega^{2}
\end{equation}
where the radial $r$ coordinate is defined such that the area of a sphere is given by $A=4\pi r^{2}$ and the $t$ coordinate gives the time as measured by a static observer at $r=\infty$. Schwarzschild radius $r_{s}=2GM/c^{2}$ is defined as the radius at which the timelike metric term proportional to $dt^{2}$ vanishes.  This denotes the boundary of the black hole called the event horizon, and also serves to define the black hole's surface area $A_{\rm BH}$.  A more physical description of the horizon is given in Fig.~\ref{fig:collapse} 
where we consider the gravitational collapse of a spherical object and the effect of the resulting horizon on the causal structure of spacetime and the propagation of photons.  

Given the relation between mass and energy, $E=Mc^{2}$, the mass-dependence of the Schwarzschild radius $r_{s}$ may be used to write the energy-conservation relation for the black hole
\begin{equation}\label{eq:conservation}
dE=c^{2}dM=\frac{\kappa c^{2}}{8\pi G}dA_{\mathrm{BH}},
\end{equation}
where 
\begin{equation}\label{eq:kappa}
\kappa=\frac{c^{4}}{4GM},
\end{equation} 
is the surface gravity of the black hole: the force/mass exerted at infinity needed to keep a small test mass stationary at the horizon.  For a  black hole, the inability of light to escape beyond the event horizon out to spatial infinity, suggests that the horizon may be viewed as a uni-directional surface.  Objects can fall into a black hole and increase its mass, but a reduction in mass is impossible as nothing can escape.  This idea was used by \cite{hawking:1972} to prove that any physical process necessarily increases the surface area of a black hole $dA_{\mathrm{BH}}\ge0$.  Shortly after, it was noted by \cite{bekenstein:1973} that this increase in area bore a striking resemblance to the second law of thermodynamics: the total entropy of an isolated system does not decrease.  This suggests that Eq.~(\ref{eq:conservation}) may be recast in the form of the first law of thermodynamics $dE=TdS$ where $T$ is the temperature of the system in thermodynamic equilibrium.  Later, the description of black hole mechanics was extended to include all four thermodynamic laws \cite{bardeen:1973}: black holes are intrinsically thermodynamical objects.

Using dimensional analysis, the relationship between area and entropy may be written in terms of the relevant fundamental constants as $dA_{\mathrm{BH}}=(\lambda G\hbar/k_{\mathrm{B}}c^{3})dS_{\mathrm{BH}}$ where $\lambda$ is an undetermined dimensionless constant.  We may therefore express Eq.~(\ref{eq:conservation}) as the thermodynamic relation
\begin{equation}
dE=T_{\mathrm{H}}dS_{\mathrm{BH}}=\frac{\hbar\kappa}{8\pi k_{\mathrm{B}}c}\lambda dS_{\mathrm{BH}},
\end{equation}
which suggests that a black hole not only absorbs energy, but also emits radiation with a temperature proportional to the surface gravity Eq.~(\ref{eq:kappa}).  This result is further motivated by the fact that the surface gravity is constant over the horizon of a stationary black hole, a property that is reminiscent of the uniform temperature of a thermal body in equilibrium; this constitutes the zeroth law of black hole mechanics \cite{bardeen:1973}.  Although these considerations argued for the existence of a black hole temperature, the inability of anything to escape beyond the horizon suggested that the effective temperature of a black hole is actually zero: $T_{\mathrm{H}}$ has no meaning as a physical temperature.  This conventional viewpoint was overturned by Hawking using quantum field theory in curved spacetime (QFTCS) to show that a black hole emits black body radiation with a Hawking temperature 
\begin{equation}\label{eq:th}
T_{\mathrm{H}}=\frac{\hbar\gamma}{2\pi k_{\mathrm{B}}}, 
\end{equation}
characterized by the surface gravity parameter $\gamma=\kappa/c$ \cite{hawking:1974,hawking:1975}.  In this way, Hawking was not only able to give a physical interpretation to the black hole temperature $T_{\mathrm{H}}$, but was also able to solidify the link between the black hole area $dA_{\mathrm{BH}}$ and entropy $dS_{\mathrm{BH}}$, with the proportionality constant fixed to be $\lambda=4$.

\begin{figure}[t]\begin{center}
\includegraphics[width=8.0cm]{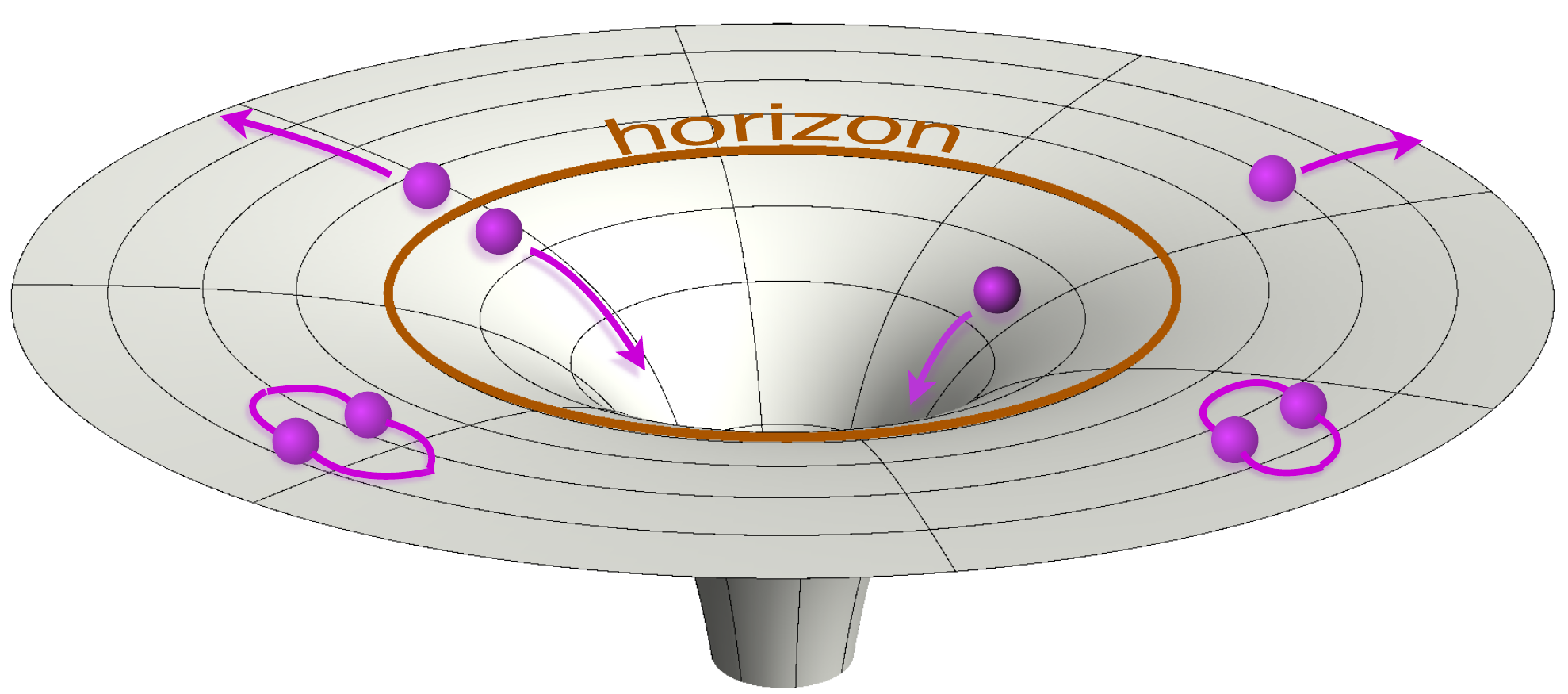}
\caption{(Color online) Cartoon of a black hole with vacuum fluctuations.  Far from the horizon, vacuum fluctuations result in virtual particles that quickly annihilate each other.  At the horizon however, one particle in a virtual pair may be trapped inside the horizon, allowing its partner to escape to arbitrary large distances---the Hawking effect.}
\label{fig:hawking}
\end{center}
\end{figure}

When viewed as a particle production process, Hawking radiation (HR) has a simple interpretation.  As shown in Fig.~\ref{fig:hawking}, vacuum fluctuations produce pairs of virtual particles that quickly annihilate each other when far from the horizon.  In contrast, near the horizon one particle in the pair may be trapped inside the horizon, unable to recombine with its partner.  The particle outside the horizon is then free to propagate out to an observer at spatial infinity.  The energy necessary for the outflow of particles comes from the gravitational field produced by the black hole's mass $M$ which, due to energy conservation, must decrease over time as radiation is emitted.  With the surface gravity (\ref{eq:kappa}) being inversely proportional to the black hole mass and proportional to the Hawking temperature, the latter increases as the black hole radiates away energy.  Unabated, the black hole experiences an unbounded increase in its temperature, and ultimately complete evaporation.

Although a black hole's mass $M$ decreases as HR is emitted, in typical derivations of the Hawking effect that use QFTCS \cite{hawking:1975,boulware:1976,hartle:1976}, the black hole mass, and therefore the spacetime metric (\ref{eq:Schwarzschild}), is considered to be fixed throughout the calculation.  This is for two reasons: (i) The power output from the Hawking process is exceedingly low for black holes with masses above the Planck mass $m_{\mathrm{P}}=\sqrt{\hbar c/G}\sim2\times 10^{-8}~\mathrm{kg}$.  In this situation, the net loss of energy due to HR is a negligibly small portion of the total black hole energy, and can safely be ignored.  For example, a relatively small black hole may be close to the mass of the sun $\sim 10^{38}~\mathrm{kg}$, and is therefore well above this Planck scale.  (ii) Allowing for black hole evaporation introduces explicit time-dependence in the spacetime metric.  However, the connection between the zeroth and first-law of thermodynamics to those of black hole mechanics relies on the assumption of a stationary spacetime and a well defined surface gravity; conditions which are violated during evaporation \cite{wald:2001}.

In essence, the fixed mass condition assumes a classical source of energy with fixed amplitude that cannot be depleted through the emission process.  Although this assumption appears to be unique to black holes, we have in fact made use of similar approximations for both the PA and UE considered in Secs.~(\ref{sec:paramp}) and (\ref{sec:unruh}), respectively.  For the PA, our use of a classical fixed amplitude pump mode plays an analogous role to the fixed black hole mass.  Likewise, in the UE we implicitly assumed that the source of the observer's acceleration had an unlimited supply of energy so as to maintain the proper acceleration $a$ indefinitely.  We can in fact make use of this fixed mass condition, via the surface gravity (\ref{eq:kappa}), to relate the emission of HR to the UE through Einstein's equivalence principle relating inertial and gravitational accelerations \cite{einstein:1907}, as we now demonstrate below.  

With HR generated close to the black hole horizon [see Fig.~\ref{fig:hawking}], the relationship to the UE is elucidated by taking the near-horizon approximation to the Schwarzschild metric Eq.~(\ref{eq:Schwarzschild}).  To explore the near-horizon region of the black hole, we replace the Schwarzschild radial coordinate $r$ with a length 
\begin{equation}
x=\int_{r_{s}}^{r}\sqrt{g_{rr}(r')}dr'=\int_{r_{s}}^{r}\left(1-\frac{r_{s}}{r'}\right)^{-1/2}dr',
\end{equation}
characterizing the proper distance close to the horizon.  Near the horizon, $x\approx 2\sqrt{r_{s}(r-r_{s})}$, and the near-horizon form of the Schwarzschild metric (\ref{eq:Schwarzschild}) expressed in terms of this proper distance becomes \cite{fabbri:2005}
\begin{equation}\label{eq:near}
ds^{2}=-\left(\gamma x\right)^{2}dt^{2}+dx^{2},
\end{equation}
where we have ignored the coordinates transverse to the radial direction, as close to $100\%$ of the HR is emitted in the lowest, $l=1$, angular momentum state \cite{page:1976}; the black hole emits as close to radially as possible \cite{bekenstein:2002}.  This is due to conformal symmetry in the near-horizon region \cite{carlip:2007}, and allows for the complete description of HR using only a single spatial dimension.  The power emitted by HR in the radial direction may then be calculated assuming the unidirectional emission of power $\dot{E}_{1\mathrm{D}}$ from a one-dimensional blackbody \cite{nation:2010}:
\begin{equation}\label{eq:hpower}
\dot{E}_{1\mathrm{D}}=\frac{\pi k_{\mathrm{B}}^{2}}{12\hbar}T_{\mathrm{H}}^{2}.
\end{equation}  

The near-horizon approximation to the Schwarzschild metric (\ref{eq:near}) is of the same form as the Rindler spacetime (\ref{eq:rindler}) of an accelerating observer, where the effective acceleration is provided by the surface gravity of the black hole $\kappa$ (\ref{eq:kappa}).  The replacement $\alpha\rightarrow\gamma$ in Eq.~(\ref{eq:rindler}), which gives the metric (\ref{eq:near}), is a manifestation of Einstein's equivalence principle, and allows us to carry over the results obtained for the UE to the present case of HR.  In particular, we can replace the acceleration parameter $\alpha$ with $\gamma$ in the Unruh temperature (\ref{eq:tu}), which then agrees with Eq.~(\ref{eq:th}) for the temperature of a black hole.  Finally, as in the UE (\ref{eq:unruh-state}) and parametric amplification (\ref{eq:par-amp-wave-function}), the photon pairs generated via the Hawking process in this near-horizon region are entangled as a two-mode squeezed state.

It should be noted however that the Hawking radiation temperature (\ref{eq:th}) applies to an observer at rest far from the black hole.  This is indicated by the use of the Schwarzschild time $t$ in (\ref{eq:near}) rather than the proper time $\tau$ of an Unruh observer from Eq.~(\ref{eq:rindler}).  The surface gravity $\kappa$ is defined with respect to the observer at infinity as
\begin{equation}\label{eq:va}
\kappa=\left.Va\right|_{r=r_{s}},
\end{equation}
where 
\begin{equation}\label{eq:accel}
a(r)=\frac{GM}{r^{2}\sqrt{1-\frac{r_{s}}{r}}}
\end{equation}
is the radial acceleration needed to keep an observer stationary at the radius $r$, and $V(r)=\sqrt{1-r_{s}/r}$ is the red-shift factor accounting for the energy lost by an escaping photon due to the gravitational potential of the black hole.  It is easy to check that Eq.~(\ref{eq:va}) agrees with our earlier definition (\ref{eq:kappa}).  We may calculate the Hawking temperature at an arbitrary radius $r$ away from the horizon taking into account the red-shift as
\begin{equation}
T(r)=\frac{\hbar(\kappa/c)}{2\pi k_{\mathrm{B}}V(r)}
\end{equation}
which, as one approaches the horizon, gives $T\rightarrow\hbar(a/c)/(2\pi k_{\mathrm{B}})$ with $a$ given by Eq.~(\ref{eq:accel}). This result is exactly the same as that obtained for the Unruh temperature (\ref{eq:tu}) in Sec.~\ref{sec:unruh}.  By removing the effects of the gravitational red-shift, HR is seen to be nothing other than the UE for an accelerating observer near the horizon.  Keep in mind that the acceleration Eq.~(\ref{eq:accel}), like the corresponding  Unruh acceleration $a$, diverges as one approaches the horizon.  Thus we establish the connection between the Unruh and Hawking effects through the equivalence principle, as summarized in Fig.~(\ref{fig:relations}). 

Even though HR has been derived in a variety of ways \cite{hawking:1975,boulware:1976,hartle:1976,parentani:2000,parikh:2000}, there remain several unanswered questions.  One concerns the trans-Planckian problem \cite{jacobson:1991,unruh:2005}, where the usual derivation of the thermal HR requires that the photon's linear dispersion relation holds up to arbitrarily high energies; classical notions of spacetime are expected to break down near the Planck energy, $E_{\mathrm{P}}=\sqrt{\hbar c^{5}/G}\sim 10^{19}~\mathrm{GeV}$.  Another problem concerns the consequences of complete evaporation of a black hole via the emission of thermal HR; information stored in the black hole is destroyed, signaling a breakdown in the unitary evolution in quantum mechanics. This is known as the information loss paradox \cite{mathur:2009}.  A third problem is the difficulty in measuring and verifying the negligibly low radiation temperatures predicted for astronomical black holes, i.e.  $T_{\mathrm{H}}\sim10^{-9}~\mathrm{K}$ for a solar mass black hole. These difficulties have called into question some of the approximations made in QFTCS calculations of HR, as well as any hope of experimental confirmation.  However, light may be shed on some of these problems by considering analogue condensed matter systems.

In preparation for discussing these HR analogues in Sec.~\ref{sec:analogue-hawking} below, we note that, from a calculational standpoint, the Schwarzschild metric (\ref{eq:Schwarzschild}) is not ideal since it is singular at the horizon.  It is therefore beneficial to choose coordinates that remain well-behaved in the horizon region.  A particularly good choice are the Painlev\'{e}-Gullstrand coordinates \cite{painleve:1921} 
\begin{equation}\label{eq:metric}
ds^{2}=-\left[c^{2}-u\left(r\right)^{2}\right]d\tau^{2}+2u\left(r\right)drd\tau+dr^{2}+r^{2}d\Omega^{2},
\end{equation}
where the Schwarzschild time $t$ is replaced by the proper time $\tau$ of a free-falling observer, while the spatial coordinate remains the same as for the Schwarzschild metric.  For an unlucky observer starting from rest at spatial infinity and free-falling into a black hole, the horizon occurs where the observer's proper time velocity $u(r)$ is equal to the vacuum speed of light $c$.

\subsection{The dynamical Casimir effect}\label{sec:dce}

\begin{figure}[t]
\includegraphics[width=8.0cm]{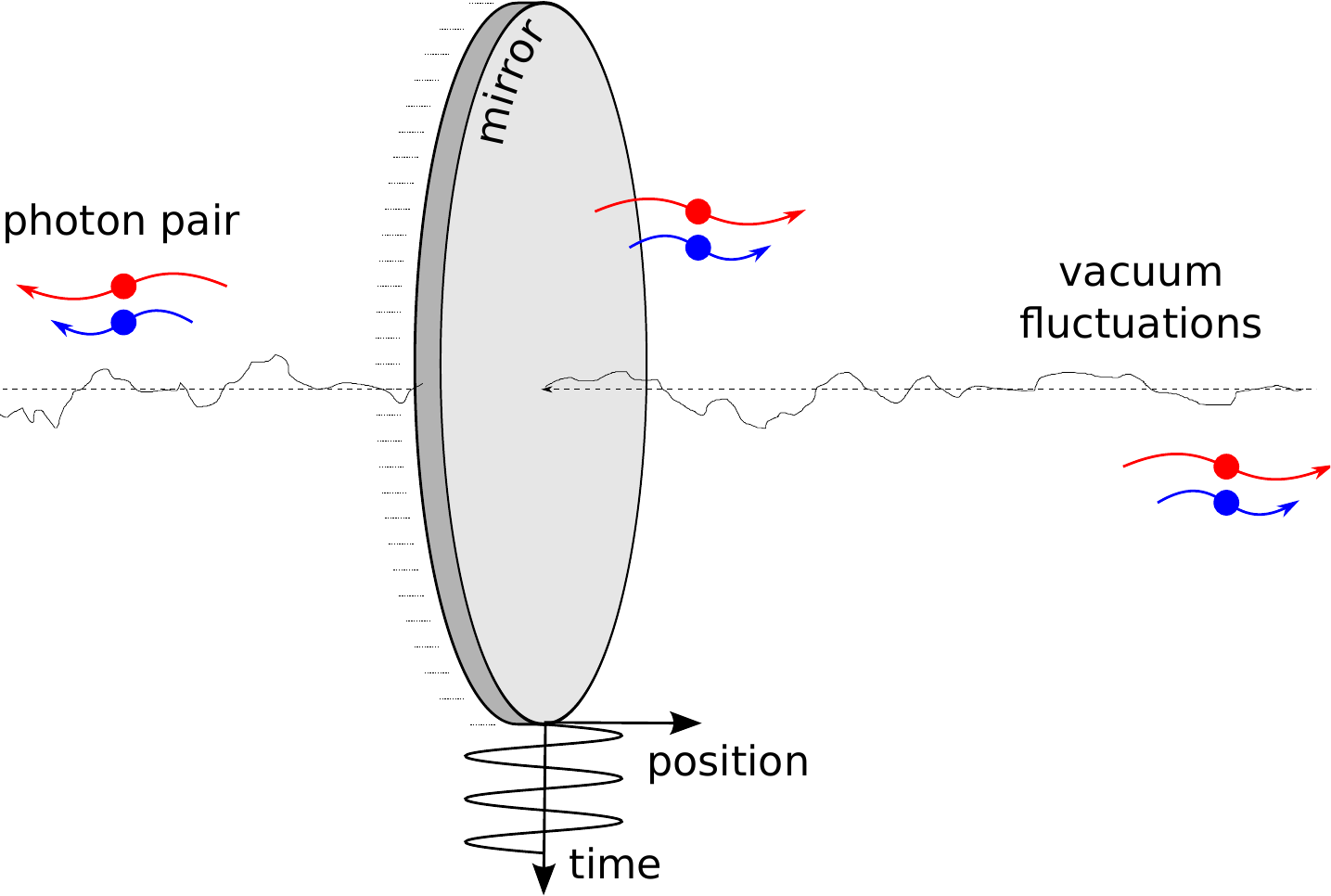}
\caption{(Color online) An oscillating mirror in free space generates photons
due to its interaction with vacuum fluctuations.
This effect is known as the dynamical Casimir effect.
The photons are generated in pairs with frequencies that add up to the frequency
of the mirror's oscillation.
The photon pair production can be interpreted as up-conversion of virtual
photons of the quantum vacuum fluctuations, 
or, equivalently, as down-conversion of pump phonons from the oscillatory motion
of the mirror.}
\label{fig:dce-schematic}
\end{figure}

The dynamical Casimir effect (DCE) concerns the generation of photons from the quantum
vacuum due to a time-dependent boundary condition, imposed by e.g.~a moving
mirror.
In contrast to the previously discussed UE in Sec.~\ref{sec:unruh}, where it was shown that
the notion of particle is observer dependent,
and where the Minkowski vacuum appears as thermal radiation to an {\it
accelerated observer},
here we will see that an {\it accelerated mirror} can result in radiation that
is detectable by an {\it inertial observer}, 
e.g., an observer at rest in Minkowski space far from the moving mirror.
See Fig.~\ref{fig:dce-schematic} for a schematic illustration of this process.

\begin{figure}[t]
\includegraphics[width=8.0cm]{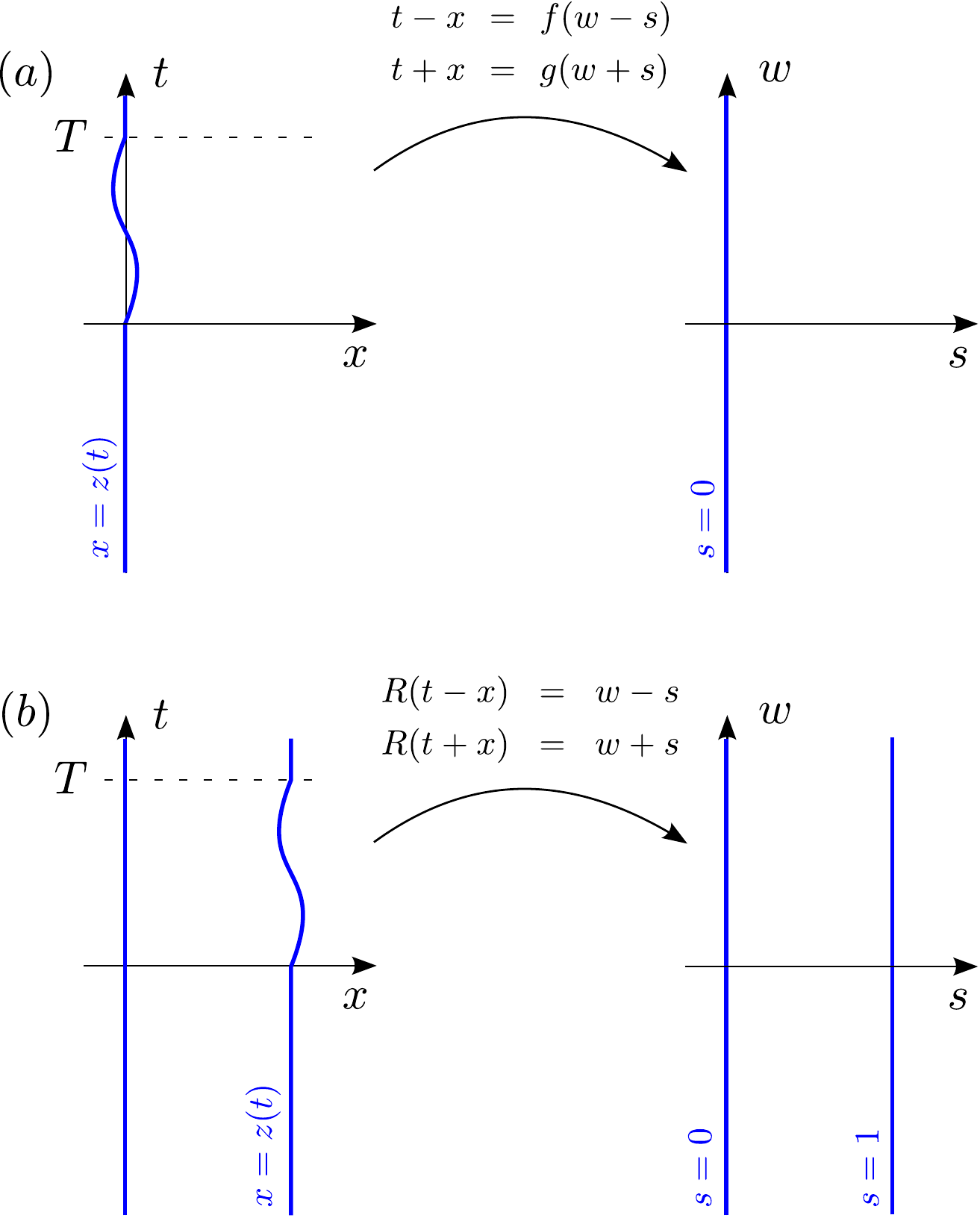}
\caption{(Color online) Mirror trajectories in the original coordinates and
the transformed coordinates for a single mirror (a) and a cavity with variable
length (b). The coordinate transformations simplify the boundary-value
problem, but finding the correct transformation functions ($f$, $g$, and
$R$, respectively) can itself be a difficult problem.}
\label{fig:dce-trajectory}
\end{figure}

Consider a massless scalar field $\phi(x,t)$ in two-dimensional spacetime
satisfying the Klein-Gordon wave equation
\begin{eqnarray}
 \frac{\partial^2 \phi}{\partial t^2} -  \frac{\partial^2 \phi}{\partial x^2} =
0,
\end{eqnarray}
and subject to the boundary condition imposed by a mirror with the
trajectory $z(t)$, 
\begin{eqnarray}
 \phi(z(t), t) = 0.
\end{eqnarray}
Following \cite{moore:1970} and \cite{fulling:1976}, we perform a  conformal (i.e. light-cone preserving)
coordinate transformation defined by
\begin{eqnarray}
\label{eq:dce_conformal_transf_f}
  t-x &=& f(w-s),\\
\label{eq:dce_conformal_transf_g}
  t+x &=& g(w+s).
\end{eqnarray}
The wave equation and the metric are invariant under conformal coordinate
transformations and retain their usual form in the $(w,s)$ coordinates:
\begin{eqnarray}
 \frac{\partial^2 \phi}{\partial w^2} &-& \frac{\partial^2 \phi}{\partial s^2} =
0,\\
dx^2-dt^2 &=& f'(w-s) g'(w+s) \left(ds^2 - dw^2\right).
\end{eqnarray}
If we impose the condition that $x = z(t)$ is mapped to $s=0$ [see
Fig.~\ref{fig:dce-trajectory}(a)],
we get the static boundary condition in the transformed coordinates
\begin{eqnarray}
 \phi(0, w) = 0,
\end{eqnarray}
and the following constraint on the functions $f$ and $g$  
\begin{equation}
\label{eq:g_and_f_transform_eq}
  \frac{1}{2}\left[g(w) - f(w)\right] = z\left\{\frac{1}{2}\left[g(w) +
f(w)\right]\right\}.
\end{equation}

In the $(w,s)$ coordinate system, the problem is static and can be readily
solved. The standard mode functions are 
\begin{eqnarray}
 \phi_\omega(w,s) = (\pi\omega)^{-\frac{1}{2}} \sin\omega s e^{-i\omega w},
\end{eqnarray}
which, in the original $(t,x)$ coordinates, take the form
\begin{eqnarray}
\!\!\!\phi_\omega(x,t) = i(4\pi\omega)^{-\frac{1}{2}} [e^{-i\omega g^{-1}(t+x)}
- e^{-i\omega f^{-1}(t-x)}].
\end{eqnarray}
The problem of finding the appropriate mode functions is therefore reduced to
finding the functions $g$ and $f$ and their inverses, given a particular mirror
trajectory $z(t)$.
For a trajectory $z(t)$, solutions that satisfy
Eq.~(\ref{eq:g_and_f_transform_eq}) usually exist, but analytical expressions
for $f(w)$ and
$g(w)$ can be difficult to obtain.

The same approach can be used for two mirrors that form a
cavity in two-dimensional spacetime \cite{moore:1970}. Assuming that one
mirror is fixed at $x=0$ and that the second mirror follow a trajectory
$x=z(t)$, the boundary conditions are
\begin{eqnarray}
 \phi(0, t) = \phi(z(t), t) = 0.
\end{eqnarray}
Applying the conformal transformation in Eqs.~(\ref{eq:dce_conformal_transf_f}-\ref{eq:dce_conformal_transf_g}),
that maps the mirror coordinates as $x=0 \leftrightarrow s=0$ and
$x=z(t) \leftrightarrow s=1$ [see
Fig.~\ref{fig:dce-trajectory}(b)], results in the static boundary condition
\begin{eqnarray}
 \phi(s=0, w) =  \phi(s=1, w) = 0.
\end{eqnarray}
Setting $f(u)=g(u)$ and denoting $f^{-1}(u) =
R(u)$ yields the constraint
\begin{eqnarray}
\label{eq:dce_moore_equation}
 R(t+z(t)) - R(t-z(t)) &=& 2.
\end{eqnarray}
This functional equation was first derived by Moore \cite{moore:1970},
and is often called the Moore equation.
Given the solution $R(u)$ to Eq.~(\ref{eq:dce_moore_equation}), we can
write the normal modes in the original $(x,t)$ coordinate as
\begin{eqnarray}
\phi_n(x,t) = (4\pi n)^{-\frac{1}{2}} [e^{-i\pi n R(t+x)} -
e^{-i\pi n R(t-x)}].
\end{eqnarray}
Again, the difficulty of the problem has been reduced to solving the
functional equation Eq.~(\ref{eq:dce_moore_equation}).

The mode functions $\phi_n(x,t)$ are orthonormal with respect to
the Klein-Gordon inner product, and can be used in the usual
canonical quantization of the field
\begin{eqnarray}
 \phi(x,t) = \sum_n a_n \phi_n(x,t) +
a_n^\dagger \bar{\phi}_n(x,t),
\end{eqnarray}
where the creation and annihilation operators $a_n$ and $a_m^\dagger$
satisfies the usual commutation relation $[a_n,a_m^\dagger] =
\delta_{nm}$. 

The state of the field can be
characterized by e.g.~the energy-momentum tensor, $T_{\alpha\beta}(x,t)$, 
\cite{fulling:1976,law:1994}
or by the photon statistics obtained by expanding the field in the Fock
state basis \cite{dodonov:1990}.
The advantage of the energy-momentum tensor, and in particular the
energy-density component $T_{00}(x,t)$, is that it is a local quantity that 
describes the radiation at the point $(x,t)$, regardless of the behavior of the
boundary conditions at that point in time, but on the other hand it requires a
regularization procedure to yield finite results. 
In contrast, the Fock-state representation is a decomposition in global 
modes that depends on the boundary condition. The photon statistics usually
gives an intuitive picture of the field state, but with time-dependent boundary
conditions there is no well-defined Fock-state basis with a
time-translationally invariant vacuum state \cite{moore:1970,fulling:1976}.
However, it is possible to formulate a meaningful photon definition by
considering a scattering-type problem for bounded motion, with stationary
mirrors in the regions $t < 0$ and $t > T$, see Fig.~\ref{fig:dce-trajectory}.
The Fock-state basis for the stationary-mirror field can be used for the in and
out regions, corresponding to $t < 0$ and $t > T$, respectively.
We can formally write the field in the stationary regions as
\begin{eqnarray}
 \phi_{\rm in}(x,t)  &=& \sum_n\left(a_n\psi^{(0)}_n(x,t) + {\rm h.c.}\right),\\
 \phi_{\rm out}(x,t) &=& \sum_n\left(b_n\psi^{(0)}_n(x,t) + {\rm h.c.}\right),
\end{eqnarray}
where $\psi^{(0)}_n(x,t) = i(\pi n)^{-\frac{1}{2}}
\sin \omega_n x e^{-i\omega_n t}$ is the mode functions for the stationary
mirror problem with resonance frequencies $\omega_n=\pi n/z_0$ and mirror separation $z_0$.
The operators $a_n$ and $b_n$ are related through the Bogoliubov
transformation
\begin{eqnarray}
 b_m &=& \sum_n\left(a_n \alpha_{nm} + a_n^\dagger \bar{\beta}_{nm}\right).
\end{eqnarray}
The coefficients $\alpha_{nm}$ and $\beta_{nm}$ are given by projecting
the mode functions for the nonstationary region $0 \leq t \leq T$ at time
$t=T$ on the stationary mirror mode functions, using the Klein-Gordon
inner product,
\begin{eqnarray}
\label{eq:bogoliubov_dce_alpha}
 \alpha_{nm} &=&  \left<\psi_m^{(0)}(x,T), \phi_n(x,T)\right>,\\
\label{eq:bogoliubov_dce_beta}
 \beta_{nm}  &=& {\left<\psi_m^{(0)}(x,T), \bar{\phi}_n(x,T)\right>^*},
\end{eqnarray}
where we have taken $\phi_{n}(x,0) = \psi_n^{(0)}(x,0)$. For the in
and
out 
regions the photon statistics is well-defined. If, for example, the field is in
the vacuum state at $t<0$, then the final photon number in the $n$th mode at
$t > T$ is
\begin{eqnarray}
 N^{\rm out}_m = \left<b_m^\dag b_m\right>_{\rm in}= \sum_n
|\beta_{nm}|^2.
\end{eqnarray}

The condition for which $\beta_{nm} = 0$ can be found by equating
the energy flux $\left<T_{01}(x,t)\right>$ to zero. \cite{fulling:1976} showed
that the mirror trajectories that result
in a field without radiation are those with uniform acceleration (including,
of course, zero acceleration).
In contrast, mirror trajectories with non-uniform acceleration result in
radiation $\left<T_{01}(x,t)\right> \neq 0$, which in the out region $t>T$
corresponds to $\beta_{nm} \neq 0$ for some $n$ and $m$. This effect is often
called the dynamical Casimir effect.

Explicit expressions for the Bogoliubov coefficients
Eqs.~(\ref{eq:bogoliubov_dce_alpha}-\ref{eq:bogoliubov_dce_beta})
and photon number $N_m^{\rm out}$ have been evaluated
for a number of different mirror trajectories with nonuniform acceleration.
A mirror trajectory of considerable theoretical interest is the exponentially
receding mirror with a velocity that asymptotically approaches the speed of light,
\begin{eqnarray}
\label{eq:receding-mirror}
 z(t) = -t - A\exp\left(-2\kappa t\right) + B, \;\;\;t>0
\end{eqnarray}
where $A, B, \kappa > 0$ are constants and $z(t)=0$, $t\leq0$.
This particular mirror trajectory results in exponential Doppler shift and radiation with a thermal black-body spectrum, with an effective temperature that is related to how fast the mirror velocity approach the speed of light $T_{\rm eff} = \kappa/2\pi$ \cite{davies:1978}.
Furthermore, an effective horizon occurs, after which a light ray from an
observer toward the mirror will never reach and reflect off the mirror, but will
instead travel to infinity along with the mirror.
Due to the appearance of this effective horizon,  the
mathematical analysis of the radiation produced by the receding mirror is
identical to the derivation of Hawking radiation from black
holes, see Sec.~\ref{sec:hawking}.
Thus we establish the connection between the dynamical Casimir effect and Hawking radiation,
as summarized in Fig.~(\ref{fig:relations}).  

From the point of view of experimentally detecting the radiation from
a non-uniformly accelerated mirror,
the most practical class of trajectories are periodic motions, and in particular
sinusoidal motion. For example, a single
mirror in free space that performs sinusoidal oscillations
produces a constant average number of photons $N_{\rm out}$ per
oscillation period \cite{lambrecht:1996, maia-neto:1996}: $N_{\rm
out} \propto (\epsilon\omega)^2$,
where $\epsilon$ is the amplitude of oscillations and $\omega$ is the
frequency of the sinusoidal mirror trajectory.

An exact solution to Eq.~(\ref{eq:dce_moore_equation}) for a cavity
with a near-sinusoidal mirror trajectory was found in \cite{law:1994}, 
where it was shown that the energy density in a cavity
with resonantly modulated length acquires a nontrivial structure in the form of
wave packets traveling back and forth in the cavity (see also
\cite{cole:1995,dalvit:1998}).
The build-up of photons in a cavity with sinusoidally modulated length
was studied in e.g.
\cite{dodonov:1990,dodonov:1993,dodonov:1996,ji:1997,schutzhold:1998}. It
was
shown that under resonant conditions, i.e., when the mirror oscillates with a
frequency that matches twice the frequency of a cavity mode, the photon
production can be resonantly enhanced.
The cavity photon number was found to grow as $(\epsilon\omega_nt)^2$ in the
short-time limit, and that the photon production rate is 
proportional to $\epsilon\omega_n$ in the long-time limit. Here $\epsilon$ is
the amplitude of oscillations and $\omega_n$ is the frequency of the resonantly
driven mode.

The rate of photon build-up in the cavity depends not only on the motion of
the cavity mirrors, but also on the mode structure of the cavity. The modes
of the ideal cavity considered in e.g.~\cite{dodonov:1990, dodonov:1993} are
equidistant in frequency, and as a result significant intra-mode interaction
occur. If, in contrast, the cavity has only a single mode, or if
intra-mode interaction is negligible due to non-equidistant frequency spacing,
the cavity can be described as a single harmonic oscillator with time-dependent
frequency \cite{dodonov:1995, meplan:1996}. The Bogoliubov transformations
Eqs.~(\ref{eq:bogoliubov_dce_alpha}-\ref{eq:bogoliubov_dce_beta}) 
for resonant driving then coincide exactly with those for a degenerate parametric
amplifier (see Sec.~\ref{sec:paramp}), and
the photon number in the cavity is therefore $N_{\rm out}=\sinh^{2}(\eta t)$,
where the squeezing parameter in this case is $\eta t = \epsilon\omega_0t$. 
Thus we establish the connection between the dynamical Casimir effect and a degenerate parametric amplifier, as indicated in Fig.~(\ref{fig:relations}).  
This correspondence between the dynamical Casimir effect in a single-mode
cavity and parametric amplification has also been discussed in
\cite{schutzhold:2005}; \cite{dezael:2010}; \cite{johansson:2010}.

It is evident from the discussion above that for the dynamical Casimir effect to
be non-negligible the modulation must also be combined with a relatively large
amplitude $\epsilon$ and high frequency $\omega$. In fact,
the maximum speed of the boundary in a sinusoidal motion $v_{\rm max} =
\epsilon\omega$, must approach the speed of light for significant photon
production to occur \cite{lambrecht:1996}.
The DCE is therefore difficult to observe in experiments using massive mirrors \cite{braggio:2005}, since such objects cannot be accelerated to relativistic velocities in practice, and therefore produce photons only at very small rates. 
The situation is improved in a cavity setup, but an important aspect that affects the photon build-up rate in a
cavity is dissipation \cite{dodonov:1998}. Although effect of dissipation is clearly to suppress the build-up of photons, a dissipative single-mode cavity with quality factor $Q$ is still expected to be above the threshold for parametric amplification if $\epsilon\omega Q > 1$ \cite{walls:2008}. A large number of photons should accumulate in such cavities, which therefore are considered promising candidates for experimental demonstration of the DCE \cite{kim:2006}.
Nevertheless, experimental verification of the DCE in the optical regime, with real massive mirrors, has not yet been demonstrated in either cavity or single-mirror setups. As previuosly discussed, this is mainly due to experimental difficulties in modulating the position of the mirrors sufficiently strongly, and the presence of decoherence, dissipation and thermal noise.

To overcome these difficulties, several systems have been proposed recently \cite{braggio:2005,segev:2007,naylor:2009,johansson:2009, johansson:2010} that use alternative means of enforcing and modulating the boundary conditions, using effective massless mirrors.
Experimental investigations of such proposals have been ongoing for the last few years \cite{braggio:2009, wilson:2010}, and have culminated in the experimental observation of the DCE \cite{lahteenmaki:2011,wilson:2011} using a superconducting waveguide.
We discuss the DCE with superconducting circuits in more detail in Sec.~\ref{sec:sc-circuits:dce}.

\section{Implementations in superconducting circuits}\label{sec:sc-circuits}
In this section we will highlight recent work, both experimental and theoretical, on implementing the amplification methods discussed in the previous section.  The possibility to generate vacuum amplification effects in superconducting circuit devices is largely due to their use in quantum information and computation \cite{you:2005,you:2011,buluta:2011}.  There, the transfer of information must be sufficiently free from dissipation and noise so as to maintain quantum coherence, while at the same time the information should be transferred via single quanta \cite{clarke:2008}.  Similar requirements are also necessary for vacuum amplification experiments, which ideally should be free from spurious photon sources, and be sufficiently coherent such that the quantum entanglement between generated particle pairs is maintained long enough to be measured.  In superconducting circuit systems, one way to achieve these combined goals is to make use of the circuit quantum electrodynamics (Circuit QED) architecture \cite{blais:2004}, where qubits are coupled via one or more effectively one-dimensional transmission line resonators \cite{chiorescu:2004,wallraff:2004,mariantoni:2011}.  Transmission lines with quality factors exceeding $\sim 10^{6}$ have been demonstrated, corresponding to a photon that travels $10~\mathrm{km}$ before being dissipated \cite{schoelkopf:2008}.  These advances have allowed for multiple qubit \cite{majer:2007,sillianpaa:2007,dicarlo:2010,wei:2006} and photon \cite{wang:2011} entanglement using transmission lines that span distances of several millimeters, and are therefore visible to the naked eye.  In addition, the generation of single-photons on demand \cite{houck:2007}, and the possibility of strong nonlinearities at the single-photon level \cite{hoffman:2010}, open up additional possibilities for the control of photons inside these devices.  Although typical experiments involve cavity resonators, recently there has been growing interest in the use of open transmission lines \cite{astafiev:2010b,astafiev:2010,zhou:2008}, which allow for broadband frequency signals such as those generated by the Unruh, Hawking, and dynamical Casimir effects.  In the sections that follow, we will describe ways to use this open 1D circuit QED architecture to generate and detect photons from the quantum vacuum.

\subsection{Single-shot microwave photon detection}\label{sec:photon}  
In order to confirm the existence of the vacuum amplification mechanisms discussed in Sec.~\ref{sec:vacuum}, one must verify that the measured photons are indeed generated from vacuum fluctuations and not some spurious ambient emission process.  One possible technique is to exploit the correlated nature of the photon emission process through the use of coincidence detection measurements of the particle pairs.  Implicit in this verification method is the use of single-shot photon detectors.  With single-shot photon measurements, one in principle has access to all orders of the statistical correlations between emitted photons, or equivalently the density matrix, and therefore has entire knowledge of the quantum state \cite{leonhardt:2010}. In the optical frequency range, such detectors are readily available and allow for, among other things, all optical quantum computation \cite{kok:2007}, Bell inequality measurements \cite{weihs:1998}, quantum homodyne tomography \cite{smithey:1993}, quantum communication \cite{bouwmeester:1997}, and encryption protocols \cite{jennewein:2000}.  In superconducting circuits, analogous single-photon detectors have been difficult to realize in practice due to the several orders of magnitude smaller energies of microwave photons as compared with visible photons. 

In the absence of photon number detectors in the microwave regime, superconducting circuit devices have made use of linear quantum amplifiers \cite{clerk:2010} such as the high electron mobility transistor (HEMT) in measuring the quantized electromagnetic fields inside resonant cavities and transmission lines.  Placed between the circuit QED system, and the secondary classical voltage or current amplification stage, these amplifiers can provide several orders of magnitude of gain for the input signal but necessarily add at least half a quantum of zero-point noise fluctuations at the input due to the Heisenberg uncertainty principle \cite{caves:1982}.  Typically, the added noise is actually much higher than this minimum value, on the order of $10-20$ photons at $5~\mathrm{GHz}$ \cite{menzel:2010}.  In using a single-photon detector, this added noise is circumvented, since an intermediate amplification stage is not required.

Recently it has been shown that a pair of linear amplifiers is capable of resolving all of the moments for the quantum state of a microwave photon provided that one repeats the experiment many times to sufficiently average out the added noise \cite{dasilva:2010,menzel:2010}. This approach has been applied to the study of blackbody radiation from a load resistor and in the investigation of quantum noise of a beam-splitter \cite{mariantoni:2010}.  Furthermore, the anti-bunching of microwave photons in a superconducting cavity has been observed by measuring the second-order coherence function \cite{bozyigit:2010}, and complete state reconstruction of propagating microwave photons was performed via homodyne tomography \cite{eichler:2010}.  In order to obtain sufficient averaging, on the order of $10^{9}-10^{10}$ repeated measurements are required. 

Unambiguous verification of the vacuum amplification mechanisms discussed in Sec.~\ref{sec:vacuum} requires on-chip single-shot photon detectors in order to measure the correlations between individual photon pairs.  Achieving this goal in the microwave regime has been one of the long-standing challenges in superconducting quantum circuits.  The first experimentally realized device capable of single-photon detection in the microwave regime was based on a double quantum dot \cite{aguado:2000} and was used in the investigation of shot-noise from a quantum point contact \cite{gustavsson:2007}.  More recently, the use of phase-qubits \cite{clarke:2008} for single-photon detection has been proposed \cite{helmer:2009,romero:2009,peropadre:2011}, driven in part by the success of similar devices in measuring and controlling the quantum state of both superconducting microwave \cite{liu:2004,hofheinz:2008,hofheinz:2009,ansmann:2009,wang:2009} and mechanical \cite{oconnell:2010} resonators.  This work has culminated in a microwave Hanbury Brown and Twiss type experiment \cite{hanbury:1956} using a pair of phase-qubit detectors, and the observation of photon bunching from a thermal source \cite{chen:2010}.  Here, the absorption of a single photon by the phase qubit causes a transition to the excited state which readily tunnels out of the potential well and into the continuum, generating a voltage signal via the Josephson phase-voltage relation \cite{likharev:1986}.  Detection efficiencies exceeding $80\%$ were achieved, although in principle a perfect detector is possible \cite{peropadre:2011}.  An ideal single phase-qubit detector acts as a binary, or ``bucket", detector that responds to the input signal by always absorbing a single photon, regardless of the original number of photons present.  Number resolving detection can be approximated using only binary detectors by detector cascading, or ``multiplexing" \cite{leonhardt:2010}, where a single incoming mode is equally distributed over a large number of output modes followed by qubit detectors.  If the number of qubit detectors is large compared to the number of photons present in the signal,  each detector receives only a single photon on average, allowing high fidelity measurements of the photon number to be performed \cite{kok:2007}.

\subsection{SQUID based microwave parametric amplifiers}\label{sec:sc-circuits:pa}
\begin{figure*}[t]\begin{center}
\includegraphics[width=18.0cm]{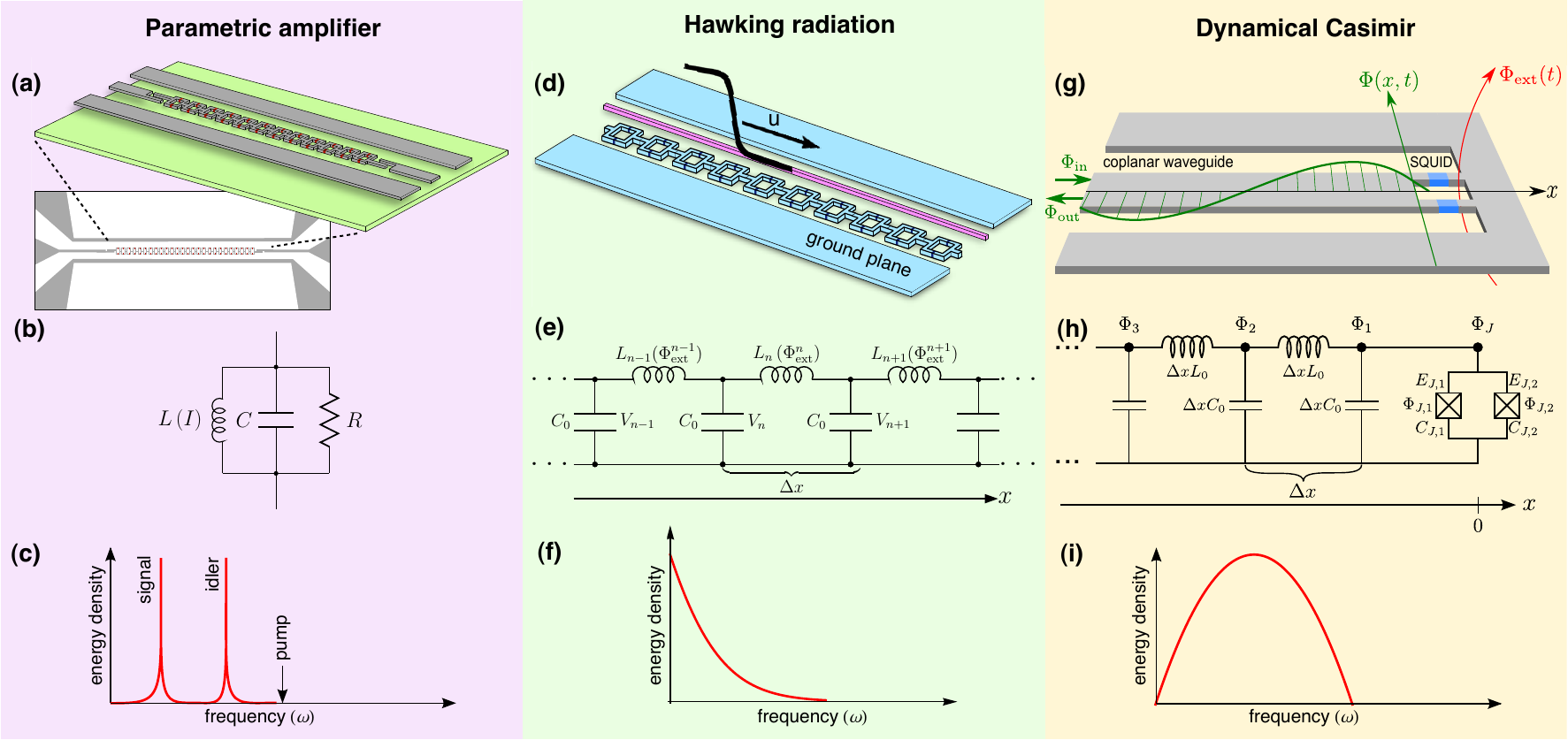}
\caption{(Color online) Schemes for superconducting circuit implementations of vacuum amplification processes: (a) The SQUID based parametric amplifier from \cite{castellanos-beltran:2008}.  (b) The parametric amplifier can be approximated as a lumped $LC$-circuit with a current dependent inductance.  The small normal current resistance is also depicted.  (c) Spectrum of a parametric amplifier.  For the non-degenerate case, one has separate peaks for the signal and idler modes satisfying $\omega_{s}+\omega_{i}=\omega_{p}$.  In contrast, the degenerate amplifier satisfies $\omega_{s}=\omega_{i}$.  (d) Illustration of a dc-SQUID array transmission line with accompanying bias line (pink) and flux-pulse (black) used in generating an analogue event horizon and Hawking radiation.  (e) Lumped circuit model valid for frequencies below the plasma frequency and negligible SQUID self inductance.  (f) One-dimensional black body spectrum of emitted Hawking radiation.  The characteristic (Hawking) temperature of the distribution is determined by the gradient of the SQUID array speed of light in a frame co-moving with the flux pulse.  (g) Circuit diagram of a SQUID terminated coplanar waveguide used in generating the dynamical Casimir effect.  Modulation of the SQUID's Josephson energy is performed by the time-varying external flux $\Phi_{\mathrm{ext}}(t)$.  (h) Equivalent lumped element circuit model for the semi-infinite coplanar waveguide and dc-SQUID.  (i) Spectrum of photons emitted by the DCE assuming the SQUID is driven by a sinusoidally varying flux.}
\label{fig:compare}
\end{center}
\end{figure*}

Parametric amplification in the microwave regime has been investigated for some time \cite{barone:1982}, with early works \cite{wahlsten:1977, yurke:1988, yurke:1989} demonstrating degenerate parametric amplification using superconducting circuits and the nonlinear properties (i.e. current-phase relations) of Josephson junctions.  The squeezing of vacuum fluctuations has also been observed \cite{movshovich:1990}.  More recently, there has been a renewed interest in these devices for amplification and frequency conversion brought on by progress in solid-state quantum metrology and information processing in the microwave regime.

Of the many examples of circuit based parametric amplifiers \cite{tholen:2007,vijay:2009}, the focus here will be on systems comprising coplanar waveguide resonators incorporating dc superconducting quantum interference devices (dc-SQUIDs).  A dc-SQUID consists of two identical Josephson junctions embedded in a superconducting loop, each with critical current $I_{c}$ and capacitance $C_{J}$ (assumed identical for simplicity).  For a negligible loop self-inductance $L\ll \Phi_{0}/2\pi I_{c}$, and large plasma frequency $\omega_{p}^{s}=\sqrt{2\pi I_{c}^{s}/(2C_{J}\Phi_{0})}$, where $\Phi_{0}=h/2e$ is the flux quantum, the SQUID behaves as a passive external flux $\Phi_{\mathrm{ext}}$ and current dependent inductor
\begin{equation}\label{eq:inductor}
L(I,\Phi_{\mathrm{ext}})=\frac{\Phi_{0}}{2\pi I_{c}^{s}}\frac{\arcsin\left(I/I_{c}^{s}\right)}{\left(I/I_{c}^{s}\right)}.
\end{equation}
Here, $I_{c}^{s}=2I_{c}\cos\left(\pi\Phi_{\mathrm{ext}}/\Phi_{0}\right)$ is the SQUIDs flux tunable critical current.  When used in a lumped-element LC-oscillator such as Fig.~\ref{fig:compare}b, the flux and current dependence of this effective inductor allows two independent ways of varying the resonance frequency of the circuit.  Just like the child on the swing in Fig.~\ref{fig:swing}, this modulation of the oscillation frequency gives rise to parametric amplification.

Systems exploiting the nonlinear response of the SQUID inductance for large input currents have been considered by 
\cite{abdo:2009} and \cite{castellanos-beltran:2007,castellanos-beltran:2008}, where the centerline conductor of the resonator contained either a single or an array of embedded SQUIDs.  These devices also make use of the inductor's flux degree-of-freedom by using a dc-bias current to introduce a controllable oscillator resonant frequency tunable over several GHz.   In \cite{abdo:2009,castellanos-beltran:2007} amplification and quadrature squeezing of an input signal was observed when operated as a degenerate amplifier and driven by a large amplitude pump mode.  Additionally, amplification and squeezing of quantum fluctuations were observed by \cite{castellanos-beltran:2008} where the use of a coplanar cavity allowed for $10~\mathrm{dB}$\footnote{A decibel (dB) is a measure of the logarithmic ratio of two powers: $L_{\rm dB}=10\log_{10}\left(P_{1}/P_{2}\right)$.  In the present case of squeezing, the powers $P_{1}$ and $P_{2}$ are given by the variances $\left(\Delta X_{1}\right)^{2}$ and $\left(\Delta X_{2}\right)^{2}$ of the quadrature operators $X_{1}$ and $X_{2}$, respectively, as defined in Sec.~\ref{sec:paramp}.} of squeezing. A diagram of this experimental setup is given in Fig.~(\ref{fig:compare}a) along with the corresponding single-mode lumped element circuit diagram in Fig.~(\ref{fig:compare}b).

The systems realized by \cite{yamamoto:2008,wilson:2010} differ from the
previous examples in their use of a SQUID operated in a linear regime with
respect to both the current and the applied magnetic flux.
In these systems the SQUID terminates a coplanar waveguide resonator,
and imposes a boundary condition that is tunable through the applied
magnetic flux. 
In addition to a dc flux bias that is used to tune the resonance
frequency, a weak ac flux modulation is applied to produce
sinusoidally time-dependent resonance frequency.
Under resonant conditions, this frequency modulation can result in parametric
amplification, and the resonator is then described, in a rotating frame,
by an effective nonlinear Hamiltonian equivalent to that of a DPA.
Modulating the flux applied through the SQUID at twice the resonance frequency
was observed to amplify a small input signal and lead to
quadrature squeezing \cite{yamamoto:2008}, and to induce 
parametric oscillations in the absence of an input signal \cite{wilson:2010}.

In addition to the longstanding work on DPA's, recently non-degenerate amplification based on a Josephson parametric converter (JPC) has been considered \cite{bergeal:2010,bergeal:2010b,bergeal:2010c}.  The setup described in \cite{bergeal:2010b} consisting of two superconducting resonators coupled to a ring of four Josephson junctions, allows for the complete separation of the signal and idler modes, both spatially and temporally.  The frequency response of such a system assuming $\omega_{p}=\omega_{s}+\omega_{i}$ is given in Fig.~(\ref{fig:compare}c).  Phase-preserving amplification with a noise level three times that of the quantum limit was demonstrated in \cite{bergeal:2010}.  Moreover, correlations between signal and idler modes of a two-mode squeezed state (\ref{eq:par-amp-wave-function}) generated from the quantum vacuum were seen in \cite{bergeal:2010c}.  These correlations have also been observed for itinerant photons generated by a non-degenerate parametric amplifier formed from a broadband transmission line resonator terminated by a SQUID \cite{eichler:2011}.  Unlike the JPC however, the use of a single resonator does not allow for spatial separation of the modes without the use of an additional beam-splitter.

\subsection{Unruh effect in driven nonlinear circuit devices}
Of the four effects considered, the Unruh effect (UE) is perhaps the most difficult to reproduce in an on-chip circuit device, since it requires the observer (two-level detector) to undergo constant acceleration;  a circuit model capable of reproducing the UE has yet to be proposed.  However, an interesting related mechanism occurs in nonlinear circuit devices driven into the bistable regime \cite{marthaler:2006,dykman:2007,serban:2007}.  Here, the emission of energy into a thermal reservoir, viewed in a coordinate system rotating at the driving frequency (i.e. the rotating frame), leads to transitions to \textit{both} higher and lower quasienergy levels \cite{dykman:2007}.   These transition rates obey a Boltzmann distribution with an effective temperature determined by the quasienergy.  Surprisingly, this effective temperature is nonzero, even when the temperature of the thermal reservoir vanishes \cite{marthaler:2006}.  This same effect was found for a two-level detector in the rotating frame \cite{serban:2007}, where a zero temperature thermal bath is seen to have both positive and negative frequency Fourier components, leading to transition rates between energy levels that are described in terms of a non-vanishing effective temperature.  These predictions have been verified experimentally using a Josephson Bifurcation amplifier\cite{vijay:2009}.  These results are similar to that of an accelerating observer in the UE, Eq.~(\ref{eq:detailed}), who views the Minkowski vacuum state as a thermal state at the Unruh temperature (\ref{eq:tu}).  Although it is tempting to consider this an analogue to the UE, the excitation of a detector in the rotating frame does not correspond to an actual thermal environment comprised of physical particles \cite{letaw:1980}.

In the UE, both the amplified vacuum state (\ref{eq:unruh-state}) and the expectation value for the number operator, derived from the Bogoliubov transformations in Eq.~(\ref{eq:unruh-bogo}), correspond to a thermal state at the Unruh temperature (\ref{eq:tu}).  However, while an observer in the rotating frame will register excitations from the vacuum as a result of negative frequency vacuum modes transforming to positive-frequency components in the rotating frame\footnote{For a discussion of this effect in nonlinear circuit devices see  \cite{serban:2007}.} \cite{letaw:1980}, the expectation value for the corresponding number operator vanishes \cite{crispino:2008}.  There is no mixing of positive and negative frequency components \cite{birrell:1982}, and no natural definition of a particle for a rotating observer  \cite{letaw:1980}.  Of course, one may still define an effective temperature for a single-mode using Eq.~(\ref{eq:detailed}), as done in \cite{serban:2007}, however in contrast to the UE, this effective temperature is frequency dependent and does not correspond to a physical thermal environment.  In Sec.~\ref{sec:unruh} we saw that the energy needed to generate particles in the UE comes from the work done by the accelerating force.  Therefore, in a rotating frame where the work vanishes, there is no particle production.  Furthermore, unlike both the UE and HR, the spacetime of an observer in circular motion does not contain a horizon \cite{letaw:1980}, the essential ingredient for generating a thermal environment of tangible particles from the quantum vacuum.

\subsection{Analogue Hawking radiation in a dc-SQUID array}\label{sec:analogue-hawking}
Observing HR in a condensed matter system was first suggested by \cite{unruh:1981} who discovered an analogy between sound waves in a fluid and a scalar field in curved spacetime.  The possibility of generating HR in a condensed matter system exists because Einstein's equations are not essential to the formal derivation of HR\footnote{This absence of Einstein's equations is a consequence of using quantum field theory in curved space which ignores back-reaction effects on the spacetime metric.  This is closely related to the classical fixed amplitude pump approximation used in the parametric amplifier of Sec.~\ref{sec:paramp}.  Although unable to reproduce the Einstein equations, analogue systems can still obtain related results when energy loss is taken into consideration \cite{anglin:1995,nation:2010a}.} \cite{visser:2003}.  Instead, HR relies on two general requirements: (i) An effective spacetime metric containing a horizon.  (ii) A quantized electromagnetic field with the correct Bogoliubov transformations for the conversion of vacuum fluctuations into photons.  Since Unruh's original proposal, analogues satisfying these conditions have been found in liquid Helium \cite{jacobson:1998}, Bose-Einstein condensates \cite{garay:2000}, electromagnetic transmission lines \cite{schutzhold:2005}, fiber-optic setups \cite{philbin:2008}, superconducting circuits \cite{nation:2009}, and ion rings \cite{horstmann:2010}.

Although a variety of systems can in principle generate HR, the requirements for the unequivocal verification of the effect are common to all setups.  First, the temperature of the emitted radiation should be higher than that of the ambient background environment so as to be detectable.  Second, one must measure the correlations across the horizon between emitted photon pairs.  This latter requirement is essential, since it is the only way to verify that a photon is emitted through the Hawking effect rather than from some other ambient emission process.  Recently, \cite{belgiorno:2010} claimed to observe analogue HR in a fiber-optical setup similar to that of \cite{philbin:2008}.  Although tantalizing, this experiment did not measure correlations between photon pairs and therefore cannot confirm the source of the emitted photons.  Other objections to this claim of analogue HR have also been raised \cite{schutzhold:2010}. Another recent experiment has also succeeded in generating stimulated Hawking emission using surface waves on water \cite{weinfurtner:2011}.  Although the spontaneous generation of particles from the Hawking effect cannot be observed in this setup, using the connection between stimulated and spontaneous emission, this work has demonstrated the thermal nature of the emission process, independent of the underlying short-wavelength physics, and the irrelevance of the full Einstein equations in the description of HR.

While not a superconducting device, the first circuit model for analogue HR was considered by \cite{schutzhold:2005}, where the horizon necessary for the conversion of vacuum fluctuations into photons was produced by modulating the capacitance of a one-dimensional (1D) microwave waveguide by means of an externally applied laser beam.  The considered waveguide was modeled as a lumped-element transmission line, where the capacitance was formed by parallel conducting plates separated by a dielectric insulting material that couples to the laser's electric field. Sweeping the laser light along the waveguide at a fixed velocity, the resulting change in capacitance in turn changes the speed of light inside the waveguide and generates a horizon.  Using experimentally feasible parameters, the Hawking temperature in this system was shown to be $\sim 10-100~\mathrm{mK}$.  These temperatures are quite promising, as they are in the range of the ambient environmental temperatures set by dilution refrigerators [ see e.g. \cite{hofheinz:2009}].

Even with these relatively large Hawking temperatures, the setup considered in \cite{schutzhold:2005} has yet to be realized in experiment.  The main drawback lies in the laser-based illumination, which would generate a large number of excess environmental photons.  Moreover, unless the waveguide is itself superconducting, heating due to dissipative processes will be a problem.  Finally, the photons in the waveguide are in the microwave regime and we therefore require a single-shot microwave detection scheme to verify the photon pair correlations.  

We have already seen how superconducting devices may be used for microwave photon detection.  We will now turn to a superconducting circuit device for the generation of analogue HR that overcomes the effects of unwanted dissipation and is based on currently available manufacturing techniques \cite{nation:2009}.  To generate analogue HR in a superconducting circuit we consider the coplanar transmission line in Fig.~(\ref{fig:compare}d), where the centerline conductor is formed from an array of dc-SQUIDs.  Additionally, a current bias line capable of applying an external flux to the SQUIDs is assumed to run the length of the array.  This setup is closely related to the DPA's in  \cite{castellanos-beltran:2007,castellanos-beltran:2008}, where we have replaced the resonator with an open transmission line in order to excite a continuum of modes.  The SQUIDs are approximated as lumped inductors (\ref{eq:inductor}), forming an LC-oscillator together with the geometric capacitance between the centerline conductor and transmission line ground planes \cite{blais:2004}, see Fig.~(\ref{fig:compare}e). Therefore, this setup is essentially an array of coupled oscillators each with a nonlinear flux-dependent frequency.  As a discrete system, our waveguide has a natural short distance, high-frequency, cutoff due to the SQUID separation $\Delta x$.  The SQUID inertial terms, ignored in the lumped inductor approximation (\ref{eq:inductor}), give an additional high-frequency scale set by the plasma frequency $\omega_{p}$.  The lowest of these two frequencies determines the onset of a nonlinear photon dispersion relation, and plays the role of the high-energy scale physics in our model \cite{unruh:2005}.  Unlike a black hole, our circuit model is well characterized at all energy scales.

\begin{figure}[t]\begin{center}
\includegraphics[width=7cm]{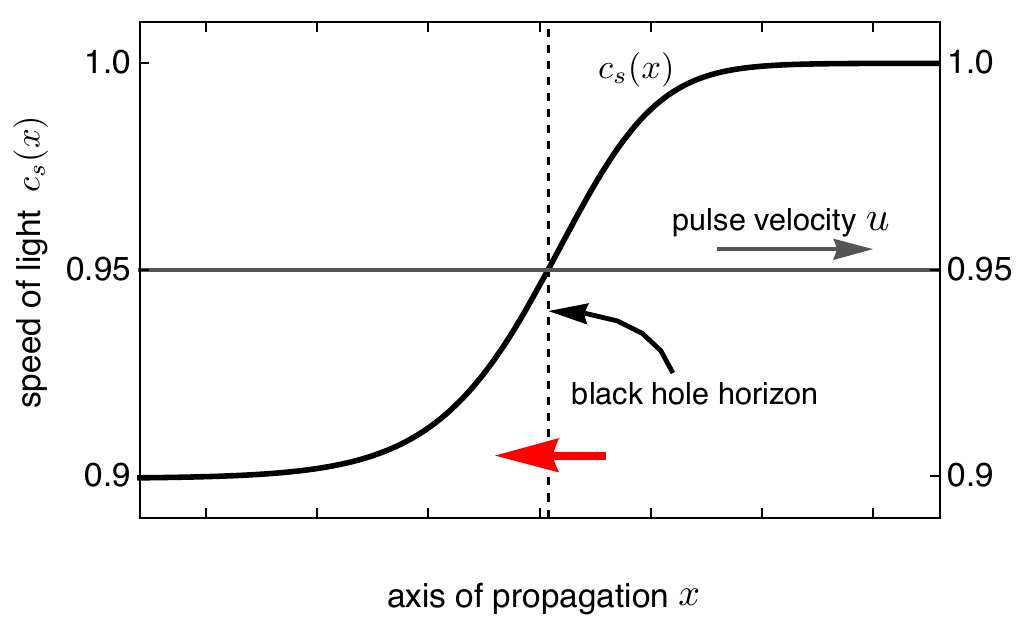}
\caption{(Color online) Effect of a steplike flux bias pulse on the SQUID array  speed of light $c_{s}(x)$ as seen in a frame moving with the pulse.  Here, velocities have been normalized with respect to the unbiased speed of light $c_{s}\left[\Phi_{\mathrm{ext}}(x)=0)\right]$.   The pulse velocity was chosen to be $u=0.95c_{s}(0)$.  In the co-moving frame, the horizon occurs where $c_{s}(x)=u$.  Like a black hole, the horizon is a unidirectional surface, and the red arrow at the bottom indicates the only permissible direction for a photon to transverse the horizon.}
\label{fig:regions}
\end{center}
\end{figure}

In order to generate the horizon, an external flux $\Phi_{\mathrm{ext}}$ is applied to the SQUID array in the form of a step-like flux pulse with fixed velocity $u$.  When the flux pulse $\Phi_{\mathrm{ext}}(x-ut)$ moves along the array, the inductance of the SQUIDs increases, resulting in a decreased speed of light in the vicinity of the pulse, 
\begin{equation}
c_{s}(x-ut)=\frac{\Delta x}{\sqrt{L\left[\Phi_{\mathrm{ext}}(x-ut)\right]C_{0}}}.
\end{equation}
Here, $L\left[\Phi_{\mathrm{ext}}(x-ut)\right]$ and $C_{0}$ are the dc-SQUID inductance and capacitance to ground, respectively.  In analogy with Eq.~(\ref{eq:metric}), the horizon is generated where the pulse velocity $u$ is equal to the SQUID array speed of light $c_{s}$.  However, recall that this definition of the horizon is valid only with respect to a moving observer.  We therefore perform a coordinate transformation into a reference frame moving with the bias pulse.  In this comoving frame, the wave equation for the electromagnetic field inside the SQUID array can be cast in terms of an effective spacetime metric with the form
\begin{equation}
ds^{2}_{\mathrm{eff}}=-\left[c_{s}\left(x\right)^{2}-u^{2}\right]d\tau^{2}+2udx d\tau+dx^{2},
\end{equation}
which is similar in form to the black hole metric (\ref{eq:metric}), apart from the interchange of spatial dependence between the SQUID array speed of light and flux pulse velocity.  In Fig.~(\ref{fig:regions}) we plot the effect of a hyperbolic tangent flux-bias pulse of amplitude $\Phi_{\mathrm{ext}}=0.2\Phi_{0}$ on the SQUID array speed of light $c_{s}$ in the comoving frame\footnote{This choice of bias-pulse is motivated in \cite{nation:2009}.}.  The pulse velocity must satisfy $u<c_{s}(\Phi_{\mathrm{ext}}=0)$ to form an horizon.

Like both HR and the UE, the analogue HR temperature is determined by the characteristic frequency of the horizon.  In condensed matter analogues, this frequency is given as the rate of change in the speed of light evaluated at the horizon 
\begin{equation}\label{eq:hawking}
T_{H}=\frac{\hbar}{2\pi k_{b}}\left|\frac{\partial c_{s}(x)}{\partial x}\right|_{c_{s}^{2}=u^{2}},
\end{equation}
resulting in a one-dimensional blackbody spectrum, Fig.~(\ref{fig:compare}f).  In addition, the output power in this device is identical to that emitted from a black hole, Eq.~(\ref{eq:hpower}).  To estimate the Hawking temperature, we will assume parameter values similar to those of the DPA in \cite{castellanos-beltran:2007}.  In addition, the validity of the SQUID inductor approximation demands that the change in the speed of light be less than the plasma frequency $\omega_{p}^{s}$.  Assuming a maximum frequency an order of magnitude smaller than the plasma frequency results in a Hawking temperature $\sim 120~\mathrm{mK}$.  This temperature can be a factor of 10 larger than the ambient temperature set by a dilution refrigerator and should be visible above the background thermal spectrum.

Unlike a real black hole, both photons in the two-mode squeezed state may be detected in this device, allowing for verification of the HR.  In the laboratory frame, a detector at the far end of the SQUID array will see two incoming photons.  One photon in front of the horizon, and one behind, with the former having a slightly higher propagation velocity (see Fig.~\ref{fig:regions}).  Single-shot detection of these microwave photons can be accomplished using one or more tunable-phase qubit detectors \cite{chen:2010} coupled to the SQUID array.  By repeatedly sending flux pulses down the bias line, the predicted one-dimensional black body spectrum may be probed by tuning the qubit resonant frequency.  Additionally, information on the cross horizon correlations between the emitted photon pairs can be established though coincidence detection.  In this way, one can unambiguously establish HR as the source of the emitted photons.

\subsection{Dynamical Casimir effect in superconducting
circuits}\label{sec:sc-circuits:dce}
Superconducting coplanar waveguides (CPWs) are excellent devices for
confining quasi one-dimensional electromagnetic fields, 
which at low (cryogenic) temperatures and GHz frequencies can behave 
quantum mechanically.
The boundary conditions for the field in a CPW can be made externally
tunable by terminating the waveguide through a SQUID.
The SQUID effectively imposes a boundary condition for the CPW, rather than being a dynamical system in itself, if its plasma frequency is much larger than all other relevant frequencies. The imposed boundary condition is then a function of the externally applied magnetic flux through the SQUID loop.
This method of implementing tunable boundary conditions has been used, e.g., in experiments on frequency-tunable resonators \cite{sandberg:2008,palacios-laloy:2008}, and for parametric amplification \cite{yamamoto:2008} and oscillations \cite{wilson:2010} (see Sec.~\ref{sec:sc-circuits:pa}). 
%
%

It has also been proposed that SQUID-terminated CPW devices can be used for experimental investigations of the DCE \cite{johansson:2009,johansson:2010}. For frequencies far below the plasma frequency, it can be shown that the boundary condition that the SQUID imposes on the CPW reduces to that of a perfectly reflecting mirror at an effective distance from the SQUID,
\begin{eqnarray}
\mathcal{L}_{\rm eff} = 
\frac{L(I, \Phi_{\rm ext})}{L_0}.
\end{eqnarray}
Here, $L(I, \Phi_{\rm ext})$ is the Josephson inductance of the SQUID [Eq.~(\ref{eq:inductor})], and $L_0$ is the characteristic inductance per
unit length of the CPW. The effective length $\mathcal{L}_{\rm eff}$ is a function of the externally applied magnetic flux $\Phi_{\rm ext}$. By applying an oscillating magnetic flux through the SQUID loop, it is therefore possible to mimic the boundary condition of an oscillating
mirror, resulting in DCE radiation.

The phase drop across a SQUID is exceptionally sensitive to the applied
magnetic flux, and the effective length of the SQUID can therefore 
be tuned in a wide range by small changes in the applied magnetic flux.
In addition, sinusoidal magnetic fields that are generated by ac
currents through bias lines adjacent to the SQUID
can reach high frequencies (tens of GHz) in state-of-the-art
experiments with superconducting circuits \cite{yamamoto:2008, wilson:2010}.
This combination of large-amplitude and high-frequency modulation makes SQUID-terminated CPWs well suited for experimental demonstration of the DCE, as this allows relatively large photon production rates. Estimates suggests that with realistic circuit parameters radiation energies on the order of mK in temperature units can be achieved \cite{johansson:2009}, which is within the limit of sensitivity in recent experiments using linear amplifiers. 

After decades of eluding experimental observation, the dynamical Casimir effect was recently demonstrated experimentally \cite{dalvit:2011,wilson:2011} using the kind of SQUID-terminated CPW device described above. In the experimental demonstration it was shown that the modulation of the boundary condition imposed by the SQUID does indeed result in photon production, and furthermore, that the generated radiation exhibits strong two-mode squeezing, which is a distinct signature of the quantum mechanical photon-pair creation process of the dynamical Casimir effect. 

Shortly thereafter, the DCE in a resonator with time-dependent dielectric properties was also demonstrated in a SQUID-array resonator \cite{lahteenmaki:2011}, similar to those used in \cite{castellanos-beltran:2007, castellanos-beltran:2008}, where the array was operated in the linear regime with a high-frequency magnetic flux field applied (uniformly) across the SQUID array. The modulation of the inductances of the SQUIDs due to the applied magnetic flux then results in time-dependent dielectric properties of the SQUID-array resonator that corresponds to a modulation of the effective length of the resonator $\mathcal{L}_{\rm eff}(t) = \mathcal{L}\sqrt{L(0)/L(t)}$, where $L(t) = L(I, \Phi_{\rm ext}(t))$ now is the characteristic inductance per unit length of the SQUID array, and $\mathcal{L}$ is the length of the resonator.  

Another type of superconducting device for studying the DCE experimentally was
introduced by \cite{segev:2007}.
That device consists of a superconducting stripline resonator that is
illuminated with an optical laser. 
The optical radiation modulates the ratio of superconducting to normal electrons
in the 
microwave stripline resonator, which in turn modulates its dielectric
properties.
Since a medium with time-dependent dielectric properties has a similar effect
on the electromagnetic field as a time-dependent boundary condition
\cite{yablonovitch:1988, johnston:1995},
it is expected that the laser illumination of the stripline resonator results
in 
photon creation due to the DCE. Promising initial experimental results for this
system has been reported \cite{segev:2007}, where a resonance frequency
shift due to the laser illumination was demonstrated.

An alternative approach to amplification of vacuum fluctuations in a superconducting circuit was proposed in \cite{deliberato:2009}. There, it was shown that a non-adiabatic modulation of the vacuum Rabi frequency (i.e., the coupling strength) in a superconducting qubit-resonator circuit can produce a significant amount of radiation. Furthermore, the resulting radiation has spectral properties that should distinguish it from spurious photon sources, such as e.g.~ambient thermal radiation.

Using CPWs or stripline resonators in experiments on the DCE has the advantage that the electromagnetic field is quasi one-dimensional. Although the general setting of the DCE is the three-dimensional free space, most theoretical work on the DCE is, for simplicity, restricted to systems with only one spatial dimension. The CPW and stripline geometries are examples of physical realizations of such systems. The fact that the photons are confined to the CPW should also simplify the process of detecting the generated radiation.
Once DCE radiation has been successfully generated, there are a number of characteristics in the photon statistics that can be used to distinguish it from spurious photon noise sources.
In particular, the DCE results in correlated photon pairs with two-mode quadrature squeezing and spectral properties that can be measured with standard homodyne detection techniques \cite{castellanos-beltran:2008}. In addition, recent development of single-photon detectors in the microwave regime \cite{chen:2010} has opened up the possibility to measure directly the correlations between individual DCE photon pairs in superconducting circuits.

\section{Summary and outlook}\label{sec:future}

We have reviewed several important quantum vacuum amplification effects; the Unruh effect, Hawking radiation, and the dynamical Casimir effect, and emphasized the interconnections between these effects.  In particular, we stressed the role of parametric amplification of vacuum fluctuations in these processes.  In addition, we have examined current and future experimental setups aimed at observing these effects, or their analogs, in superconducting electrical circuits.

As we have shown, superconducting circuits are very promising devices for experimental investigations of quantum vacuum amplification effects, and such circuits have already been used in the experimental demonstration of the DCE \cite{wilson:2011,lahteenmaki:2011}.  It appears likely that more such experiments will be carried out in the near future.  In fact, several promising experimental steps in this direction have been demonstrated already in a variety of systems \cite{segev:2007, castellanos-beltran:2008, yamamoto:2008, wilson:2010}.  A particularly important experimental breakthrough has been the recent development of single-photon detectors in the microwave regime \cite{chen:2010}.  Should microwave single-photon detectors become readily available, the detection of both the DCE and HR in microwave circuits would be greatly simplified.  This would allow probing of the quantum statistics for the resulting radiation so as to identify the characteristic signatures of these effects.

In addition to the quantum vacuum amplification effects discussed in this review, superconducting circuits have also been proposed for realizing systems with ultra-strong atom-cavity coupling \cite{nataf:2010,peropadre:2010,ashhab:2010}.  The cavity field in these systems can have exotic properties such as particles in the ground state, squeezing of field quadratures, and ground state entanglement between the cavity field and the atom. Moreover, the ability to create degenerate vacuum states in a qubit array \cite{nataf:2010}, allows for the possibility of vacuum state qubits and quantum computation.  Atom-cavity systems in the ultra-strong coupling have only recently started to become feasible experimentally \cite{niemczyk:2010,forn-diaz:2010}. This is yet another example of new regimes in quantum mechanics that are starting to become accessible due to progress in the engineering of quantum superconducting circuits.

Finally, as a quantum coherent device, the superconducting arrays of SQUIDs presented here may allow for investigating effects analogous to those of quantum gravitational fluctuations on the Hawking process and the propagation of photons.  Making use of the superconducting-to-insulator phase transition in the SQUID array \cite{chow:1998,haviland:2000}, the application of a sufficiently large external flux results in quantum fluctuations of the dynamical variables governing the SQUID inductance in Eq.~(\ref{eq:inductor}).  As this inductance determines the speed of light inside the array, this result may be interpreted as analogue fluctuations of the effective spacetime metric \cite{nation:2009}.  For analogue Hawking radiation, these fluctuations manifest themselves as quantum uncertainty in the position of the horizon in Eq.~(\ref{eq:metric}), a scenario that is of interest for actual black holes as well \cite{ford:1997,parentani:2001}. As discussed in Sec.~\ref{sec:analogue-hawking}, our condensed matter analogues cannot faithfully reproduce the full Einstein equations, and the effective metric fluctuations do not provide an analogue of the yet to be determined dynamics expected from the quantum theory of gravity [e.g. the Wheeler-Dewitt equation \cite{dewitt:1967}].  Nevertheless, given that a theory of quantized gravity remains out of reach for the foreseeable future, the ability to reproduce analogous fluctuating metric effects in a superconducting circuit model should prove useful in addressing quantum gravitational corrections to the Hawking effect.

Given the ability to fabricate a wide range of devices, the full scope of quantum vacuum effects in superconducting circuits, and the possible applications thereof, is still unknown and in need of further investigation.  Indeed, the superconducting circuit models discussed here are an example of quantum simulators \cite{lloyd:1996,buluta:2009}: controllable quantum systems engineered to reproduce the physical properties of another, seemingly different, quantum system.  The wide range of amplification effects that can be simulated in these systems, hints at the possibility of a circuit-based universal quantum vacuum amplification simulator; a device capable of exploiting the generality of Bogoliubov transformations to reproduce the emission properties of any vacuum amplifier.  What is certain however, is that superconducting circuits as a test bed for quantum-vacuum related physics offer unique advantages that will help to shed light on one of quantum mechanics' most remarkable features, namely the amplification of vacuum quantum fluctuations.

\section*{Acknowledgements}
We thank the referees for their very helpful comments on this Colloquium.  PDN was partially supported by the Japanese Society for the Promotion of Science (JSPS) Postdoctoral Fellowship No.~P11202.  MPB acknowledges support by the National Science Foundation (NSF) under grant No.~DMR-0804477.  FN was partially supported by DARPA, AFOSR, Laboratory for Physical Science, National Security Agency, Army Research Office, NSF grant No. 0726909, JSPS-RFBR contract No. 09-02-92114, Grant-in-Aid for Scientific Research (S), MEXT Kakenhi on Quantum Cybernetics, and Funding Program for Innovative R\&D on S\&T (FIRST).

\bibliography{references}
\end{document}